\documentclass[usenatbib]{mn2e}
\usepackage{bm}
\usepackage{amssymb}
\usepackage{graphicx}
\usepackage{amsmath}
\usepackage{journals}

\newcommand{\ppd}{protoplanetary disc}
\newcommand{\mri}{magnetorotational instability}
\newcommand{\half}{{\textstyle \frac{1}{2}}}
\newcommand{\thalf}{\textstyle \frac{3}{2}}

\newcommand{\cross}{\bm{\times}}
\newcommand{\etaH}{\eta_\mathrm{H}}
\newcommand{\etaA}{\eta_\mathrm{A}}
\newcommand{\etaP}{\eta_\mathrm{P}}
\newcommand{\etaO}{\eta}

\renewcommand{\div}{\grad \cdot}
\newcommand{\curl}{\grad \cross}
\newcommand{\grad}{\nabla}

\newcommand{\vv}{\bm{v}}
\newcommand{\vvk}{\bm{v}_{\rm K}}
\newcommand{\J}{\bm{J}}
\newcommand{\B}{\bm{B}}
\newcommand{\E}{\bm{E}}            
\newcommand{\dv}{\bm{\delta\!v}}
\newcommand{\dB}{\bm{\delta\!B}}
\newcommand\mat[4]{\left(\begin{array}{rr} #1 & #2 \\ #3 & #4 \end{array}\right)}
\newcommand{\Delt}[1]{\frac{\partial #1}{\partial t} + 
\Omega\frac{\partial #1}{\partial \phi} }
\newcommand{\delt}[1]{\frac{\partial #1}{\partial t}}

\newcommand{\Bh}{\hat{\bm{B}}}
\newcommand{\phih}{\bmath{\hat{\phi}}}  
\newcommand{\rh}{\bmath{\hat{r}}}     
\newcommand{\zh}{\bmath{\hat{z}}}     

\newcommand{\gpersc}{\,\mathrm{g\,cm^{-2}}}

\title{Hall diffusion and the magnetorotational instability in protoplanetary discs}
\author[Mark Wardle \& Raquel Salmeron]
{Mark Wardle$^1$ \& Raquel Salmeron$^2$\\
$^1$Department of Physics \& Astronomy and Research Centre for Astronomy, Astrophysics \& Astrophotonics,\\ Macquarie University, Sydney NSW 2109,
Australia\\
$^2$Planetary Science Institute, Research School of Astronomy \& Astrophysics and 
Research School of Earth Sciences, \\
Australian National University, Canberra ACT 2611, Australia}

\date{\today}
\pagerange{\pageref{firstpage}--\pageref{lastpage}}
\pubyear{2011}
\begin{document}
\maketitle
\label{firstpage}
\begin{abstract}
The destabilising effect of Hall diffusion in a weakly-ionised,  Keplerian disc allows the magnetorotational instability (MRI) to occur for much lower ionisation levels than would otherwise be possible.  However, simulations incorporating Hall and Ohm diffusion give the impression that the consequences of this for the non-linear saturated state are not as significant as suggested by the linear instability.  Close inspection reveals that this is not actually the case as the simulations have not yet probed the Hall-dominated regime.  Here we revisit the effect of Hall diffusion on the magnetorotational instability and the implications for the extent of MHD turbulence in protoplanetary discs, where Hall diffusion dominates over a large range of radii.  

We conduct a local, linear analysis of the instability for a vertical, weak magnetic field subject to axisymmetric perturbations with a purely vertical wave vector.  In contrast to previous analyses, we express the departure from ideal MHD in terms of Hall and Pedersen diffusivities $\etaH$, and $\etaP$, which provide transparent notation that is directly connected to the induction equation.  This allows us to present a crisp overview of the dependence of the instability on magnetic diffusivity.  We present analytic expressions and contours in the $\etaH$--$\etaP$ plane for the maximum growth rate and corresponding wave number, the upper cut-off for unstable wave numbers, and the loci that divide the plane into regions of different characteristic behaviour.  We find that for $\mathrm{sign}(B_z)\etaH<-2 v_A^2/\Omega$, where $v_A$ is the Alfv\'en speeds and $\Omega$ is the Keplerian frequency, Hall  diffusion suppresses the MRI irrespective of the value of $\etaP$. 

In the highly-diffusive limit the magnetic field decouples from the fluid perturbations and simply diffuses in the background Keplerian shear flow.  The diffusive MRI reduces to a diffusive plane-parallel shear instability with effective shear rate  $\thalf\Omega$.  We give simple analytic expressions for the growth rate and wave number of the most unstable mode.

We review the varied and confusing parameterisations of magnetic diffusion in discs that have appeared in the literature, and confirm that  simulations examining the saturation of the instability under Hall-Ohm diffusion are consistent with the linear analysis and have yet to probe the ``deep'' Hall regime  $|\etaH| > \etaP > v_A^2/\Omega$ characteristic of protoplanetary discs where Hall diffusion is expected to overcome resistive damping.

Finally, we illustrate the critical effect of Hall diffusion on the extent of dead zones in \ppd s by applying a local stability criterion to a simple model of the minimum-mass solar nebula at 1\,au, including x-ray and cosmic-ray ionisation and a population of 1\,\micron\ grains.  Hall diffusion increases or decreases the MRI-active column density by an order of magnitude or more, depending on whether B is parallel or antiparallel to the rotation axis, respectively.  We conclude that existing estimates of the depth of magnetically active layers in protoplanetary discs based on damping by Ohm diffusion are likely to be wildly inaccurate.
    
\end{abstract}
\begin{keywords}
accretion, accretion discs -- instabilities -- MHD -- planetary systems: protoplanetary discs -- stars: formation
\end{keywords}

\section{Introduction}
\label{sec:intro}

The mechanism by which angular momentum is lost by the material in accretion discs is a classic problem in astrophysics.  While gravitational torques associated with a binary companion (e.g.~\citealt{SMIH87}) or gravitational instabilities in massive discs \citep{ARS89} may play an occasional role, magnetic tension appears likely to be the most common source of the torque required to remove angular momentum from the material in the disc and permit it to spiral towards the central object.

Magnetically-mediated accretion disc models fall into two broad paradigms. In disc-jet models, a large-scale poloidal field threads the disc and the magnetic tension of the field lines anchored to the disc accelerates material outward and upward along the field lines away from the disc surfaces to form a collimated bipolar jet \citep{BP82}.  The net effect is to transfer angular momentum from the disc material to the accelerated outflow.  In the absence of an outflow the disc may still be braked through torsional magnetic coupling to an extended envelope or surrounding cloud \citep[e.g.][]{KK02}.

In models invoking magnetically-driven turbulence, on the other hand, tension in a weak magnetic field virulently destabilises the orbital shear flow in the disc via the magnetorotational instability (MRI; \citealp{V59}; \citealp{C60}; \citealp{BH91}), driving MHD turbulence that transports angular momentum radially outwards (e.g.\ \citealp{BH92}; \citealp{SHGB96}).  Generally, one expects that the jet models apply when the magnetic and thermal pressures are comparable at the disc  mid-plane, while substantially sub-thermal magnetic fields are subject to the MRI and are incapable of driving strong jets.   It is conceivable that these two transport mechanisms may operate at the same radius for intermediate strength fields, with the MRI operating for a range of $z$ adjacent to the  mid-plane of the disc and vertical transport dominating at greater heights (e.g.~\citealt{SKW07}).  Similarly, as the field strength is expected to decrease with radius, there could also be an inner region where jets dominate, with the MRI operating at larger radii \citep{CF08}.

In this paper we focus on the efficacy of the MRI-driven turbulence in protostellar/protoplanetary discs.  In this context, apart from transporting angular momentum,  MRI-driven turbulence selectively heats the disc, effectively mixes chemical species
\citep[e.g.][]{SWH06}, and affects the transport, aggregation, and sedimentation of dust grains \citep{JK05, TWBY06, FP06}, the assembly blocks of planetesimals in the `core accretion' scenario of planet formation.  Furthermore, the effective viscosity of the disc associated with MHD turbulence determines the ability of protoplanets to open gaps in the disc and also contributes to the delicate imbalance of gravitational torques that determines the rate and direction of planetary migration e.g.~\citep{MP03, NP04, JGM06}.  

It is of great importance, therefore, to determine the location, extent and behaviour of the magnetically-active regions of protoplanetary discs, which are restricted by the very low fractional ionisation beyond  $\sim 0.1$ au from the central star, where thermal ionisation is ineffective \citep{G96}. The dominant sources of ionisation -- X-rays or UV radiation from the forming star, and possibly interstellar cosmic rays -- are unable to penetrate to the disc  mid-plane except perhaps at larger radii where the column density is low.   At intermediate radii a very low level of ionisation is maintained by cosmic rays or, if these are absent, the decay of radioactive elements mixed with the gas (\citealt{SWH04, GFMW05}). 
MRI--driven MHD turbulence is therefore thought to be restricted to the surface layers \citep{G96} over the range $\sim 0.3$--$20$ au where shielding of external ionisation sources results in very low levels of ionisation.  

Assessments of the location and extent of magnetically-active zones (or equivalently, their complementary magnetic `dead zones') typically invoke only damping of the linear MRI by Ohm diffusion (e.g.~\citealt{H81,J96,FTB02,MP03,IN06b,TS08}).  However, Ohm diffusion is the dominant magnetic diffusivity only at the very high densities achieved close to the disc mid-plane within 1 au of the central star.   Instead, Hall diffusion\footnote{Despite the lack of dissipation associated with Hall drift we adopt the term ``Hall diffusion'' to maintain consistency with the notion of a tensor diffusivity appearing in the induction equation.} dominates at the intermediate  mid-plane densities between 1 and 30 au \citep{WN99,SS02a}, and ambipolar diffusion beyond about 30\,au.  Note, however, that as the density decreases towards the disc surfaces, different diffusivities may dominate in different layers.  The neglect of ambipolar diffusion in assessing magnetic activity is less drastic, as it dominates near the disc surface where the fractional ionisation is relatively high and magnetic diffusivity may be insufficient to stabilise the MRI (e.g. \citealt{PC11}).

Hall diffusion provides a dissipation-free pathway for the magnetic field that either enables the MRI to grow despite the damping effect of Ohm or ambipolar diffusion or suppresses it entirely \citep{W99,BT01}, depending on the orientation of the magnetic field.   Based on the linear growth rates Hall diffusion may drastically extend or restrict the reach of magnetic activity in protoplanetary discs \citep{W07} \emph{and likely modifies the transport and dissipative properties of the resulting turbulence.}   However, pioneering shearing-box calculations including both Ohm and Hall diffusion did not appear to confirm this expectation, suggesting instead that Hall diffusion was unable to counter the damping effect of Ohm diffusion \citep{SS02a,SS02b}.  As a result, subsequent evaluations of the extent of dead zones in protoplanetary discs are based either on the damping effects of Ohm diffusion (e.g. \citealt{TS08,TD09,KL10}), or on Ohm and ambipolar diffusion \citep{PC11,B11c}.

The apparent failure of fully-developed turbulence to respect the underlying linear instability in the Hall-diffusion-dominated limit is surprising, as it does so in the Ohm (e.g.~\citealt{TS08}) and ambipolar diffusion regimes \citep{BS11}.  However, closer inspection of the parametrisation of the diffusive effects adopted by \citet{SS02a,SS02b} reveals that their simulations did not probe the important regime in which (a) Hall diffusion dominates Ohm diffusion \emph{and} (b) both dominate inductive effects.  Indeed, none of the results of \citet{SS02a,SS02b} conflict with expectations based on the linear analysis.  The confusion that has arisen can largely be attributed to the adoption of a magnetic Reynolds number $Re_M$ and a Hall parameter $X$ to characterise the magnitudes of Ohm and Hall diffusion respectively.  Then the criteria for Ohm and Hall  diffusion to dominate the inductive term are $Re_M<1$ and $X>2$, respectively.  Crucially, the ratio of Hall to Ohm diffusion is then $\frac{1}{2} XRe_M$.  This is never $\gg 1$ when $Re_M<1$ in the simulations, apart from the zero-net-flux simulations S14--S16 of \cite{SS02b} in which Hall diffusion suppresses the MRI in the $B_z<0$ regions, as one might expect based on the linear analysis.  

The purpose of this paper is to emphasise the importance of Hall diffusion for MRI-driven turbulence in protoplanetary discs.   We do this by considering the simplest example of the MRI -- perturbations with wave vectors parallel to an initially vertical magnetic field -- and demonstrating that the column density of the MRI-unstable region at 1\,au in the minimum mass solar nebula increases or decreases (depending on the sign of $B_z$) by an order of magnitude when Hall diffusion is included  (see Fig. \ref{fig:active-sigma}).  We begin by describing  the field-line drifts induced by  ambipolar, Hall, and Ohm diffusion in \S2 and give a qualitative discussion of their effect on the MRI.  In \S3 we present an overview of the results of a linear analysis of the MRI in the presence of diffusion and give a coherent survey of its dependence on Pedersen (ie Ohm+ambipolar) and Hall diffusion by considering contours of the growth rate and wave number of the fastest growing mode and the range of MRI-unstable wave numbers in \S4.  We then compare previous notation with the diffusivity notation we now favour and discuss the numerical simulations of \citet{SS02a,SS02b}.  In \S5 we estimate the column density of the MRI-unstable layer at 1\,au in the MMSN using the criterion that there should exist a local unstable MRI mode with wave number satisfying $kh>1$.  The results are dramatic, indicating that Hall diffusion increases the column density of the MRI-unstable surface layers by an order of magnitude depending on whether $B_z$ is positive or negative, respectively. Finally, we summarise our results and conclusions in \S6.

\section{Magnetic diffusion and the magnetorotational instability}
\label{sec:diffusion}
\subsection{Magnetic diffusion}
The magnetic field evolves according to the induction equation
\begin{equation}
\delt{\B} =  \curl(\vv\cross\B) - c\curl\E'
\,.
    \label{eq:induction}
\end{equation}
where $\E'$ is the electric field in the local instantaneous rest frame of the fluid, which satisfies the generalised Ohm's Law
\begin{equation}
c \E' = \etaA (\curl\B)_\perp + \etaH(\curl\B)\cross\Bh+ \etaO(\curl\B)\,,
    \label{eq:E-fluid-P}
\end{equation}
where we use subscripts $\parallel$ and $\perp$ to refer to the orientation with respect to the local magnetic field, and $\etaA$, $\etaH$, and $\etaO$ are the ambipolar, Hall, and Ohm diffusivities, respectively.  
It will prove useful to recast the induction equation in a form that makes explicit the drift of the magnetic field through the fluid, i.e.
\begin{equation}
    \delt{\B} =  
    \curl\left[(\vv+\vv_{\rm B})\cross\B \
    -\etaO(\curl\B)_\parallel\right]
    \,,
    \label{eq:induction-v2}
\end{equation}
where
\begin{eqnarray}
    \vv_{\rm B} & = & c\;\frac{\E'\cross\B }{B^2} \nonumber\\
    & =  & \etaP\,\frac{(\curl\B)_\perp\cross\Bh}{B} 
    \,-\,\etaH\,\frac{(\curl\B)_\perp}{B} \,.
    \label{eq:vB}
\end{eqnarray}
and the Pedersen diffusivity $\etaP$ is given by
\begin{equation}
     \etaP = \etaA+\etaO,
    \label{eq:etaP}
\end{equation}
This form makes explicit the role of the diffusivities in transporting flux, and emphasizes that only Ohm diffusion is capable of destroying magnetic flux (i.e.\ via the $\etaO(\curl\B)_\parallel$ term). 

The diffusivities are determined by the abundances of charged particles and their collision cross-sections with the neutrals (e.g.~\citealt{C57,WN99}), and -- when the fractional ionisation is not too small -- with each other.  In general, $\eta$, $\etaA$, and $\etaP$ are all positive, whereas $\etaH$ is positive or negative depending on whether positively- or negatively-charged species, respectively, are on average more decoupled from the magnetic field by collisions.  For example, for a weakly-ionised ion-electron-neutral plasma we obtain
\begin{equation}
    \eta = \frac{c^2 m_e\gamma_e\rho}{4\,\pi e^2 n_e}\,,
    \label{eq:eta}
\end{equation}
\begin{equation}
    \etaA = \frac{B^2}{4\,\pi\,\gamma_i\,\rho\rho_i}\,,
    \label{eq:etaA}
\end{equation}
and
\begin{equation}
    \etaH =\frac{c\,B}{4\,\pi\,e\,n_e}\,.
    \label{eq:etaH}
\end{equation}
(e.g~\citealt{K89,BT01}). 
The diffusivities exhibit more complex dependencies on density, field strength and electron abundance when charged grains are present.  In this case the Hall diffusivity is negative at relatively low densities when neutral collisions decouple grains from the magnetic field, but not the electrons and ions, and is positive at higher densities when ions are also decoupled from the magnetic field \citep[e.g.][]{WN99}. The diffusivities exhibit significant spatial gradients if the ionisation rate and gas density do so, as in protoplanetary discs (e.g.\ \citealt{W07}).  Fortunately, for a linear analysis the response of the diffusivities to perturbations is not needed as $\curl\B$ vanishes in the initial equilibrium state.  Thus in our local analysis we can regard the diffusivities as specified constants, but we shall return to them when considering the application to \ppd s in \S\ref{sec:ppds}.

\subsection{Effect on the MRI}
Before deriving the dispersion relation for the \mri\ in diffusive
media we first give an intuitive, physical description of the
effect of magnetic diffusion on its growth and properties.

\begin{figure}
    \centering
    \includegraphics[scale=0.9,trim= 0 0 65 0]{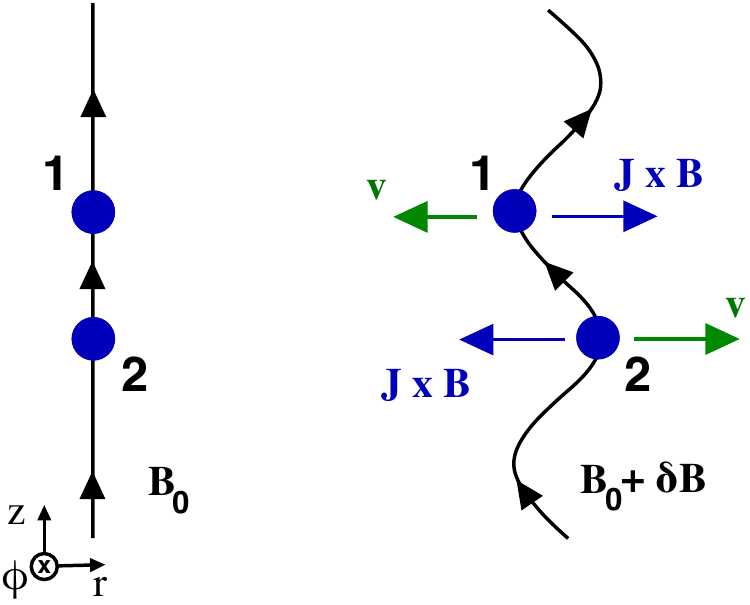}
    \caption{A sketch of the development of the magnetorotational instability (MRI) in an initially vertical magnetic field $B_0$ embedded in an unstratified, Keplerian disc for a vertical wave vector. Top panel shows a poloidal view of (\emph{left}) the initial field configuration and (\emph{right}) the perturbed field and associated current density ($\J$), Lorentz force ($\J \cross \B$) and deviation from local Keplerian velocity. The lower panel shows in plan view the perturbed fluid elements at points 1 and 2.  At point 1 the magnetic tension provides some radial support against gravity and supplies a negative torque. As a result the material here is sub- Keplerian and moving inwards, as indicated by the green vector which indicates the departure from  Keplerian rotation. The field transfers the angular momentum to the fluid at locations such as 2 where the field is buckled outwards, and the fluid there spirals out. The differential radial motion of the fluid that is losing and gaining angular momentum at 1 and 2 respectively enhances the buckling, leading to runaway growth.
    \label{fig:MRI-sketch}
    }
\end{figure}

We first examine the growth of the instability in a Keplerian accretion
disc under ideal-MHD conditions and assuming an initially vertical magnetic field (see Fig.~\ref{fig:MRI-sketch}, top-left panel). We consider, in particular, the situation where alternate layers of fluid are displaced from their equilibrium position, such that point 1 moves radially inwards (and forwards in azimuth) while point 2 is shifted radially outwards (and backwards in azimuth). In this case the magnetic field is buckled because the field lines are frozen into the fluid (top-right panel). The deformation of the lines creates magnetic tension forces which are directed outwards and backwards in azimuth (at point 1), or inwards and forwards in azimuth (at point 2), as depicted in the lower panel of Fig.\ 1. At point 1 then the magnetic tension provides some radial support against gravity and supplies a negative torque. This means that the fluid element at point 1 will lose angular momentum and spiral inwards. The opposite takes place at point 2, with the fluid element there spiralling outwards. As a result, the magnetic field buckling and associated tension increases, leading to runaway growth. The fastest growing mode has growth rate $\frac{3}{4}\,\Omega$ on scales with wave numbers $k\sim \Omega/v_{\rm A}$, where $\Omega = v_{\rm K}/r $ is the Keplerian frequency and $v_{\rm A}$ is the Alfv\'en speed \citep{BH91}.

Magnetic diffusion modifies the above picture by allowing slippage between the field lines and the fluid, so that there is no longer a direct connection between the relative displacement of the fluid layers and the buckling of the magnetic field lines.  One might expect that this effect should always reduce the instability because the displacement of the field would lag that of the fluid.  However, this intuition is based on the limit of Ohm or ambipolar diffusion, in which the drift is in the direction of the magnetic stresses and tends to straighten up field lines.  Hall diffusion, by contrast, creates a drift orthogonal to the tension forces and may, therefore, enhance or suppress the radial buckling depending on the situation.  It is this feature that gives Hall diffusion its unique properties and that we aim to explain here.

In the ambipolar diffusion limit, the magnetic field is frozen into the ions and electrons, which drift together through the neutral component of the fluid.  Collisions with the neutrals then transmit magnetic stresses to the bulk of the gas.  As the fractional ionisation is low, the ion and electron inertia and thermal pressure are negligible and the drift velocity of the ions, electrons and field is determined by the balance between the Lorentz force on the ions and the collisional drag with the neutrals,
\begin{equation}
    \vv_{\rm P} = \vv_i - \vv = \frac{\J\cross\B}{c\gamma_i\rho_i\rho}\,,
    \label{eq:ion-mom}
\end{equation}
and so is parallel to $\J\cross\B$.    In eq.  (\ref{eq:ion-mom}),
$\rho$ and $\rho_i$ are the neutral and ion density, respectively, and
\begin{equation}
\gamma_i = \frac{\langle \sigma v \rangle_i}{m_i + m} \,,
\end{equation}
where $\langle \sigma v \rangle_i$ measures the rate coefficient of momentum exchange via collisions with the neutrals, taken to have mean mass $m$, and $m_i$ is the ion mass.  When $\J\cdot\B=0$, such as envisaged here,  Ohm diffusion also produces a drift in the same direction, so we have adopted the subscript P to denote the Pedersen drift (see eq \ref{eq:vB}). 

\begin{figure}
    \centering
     \includegraphics[scale=0.8,trim= 0 0 0 0]{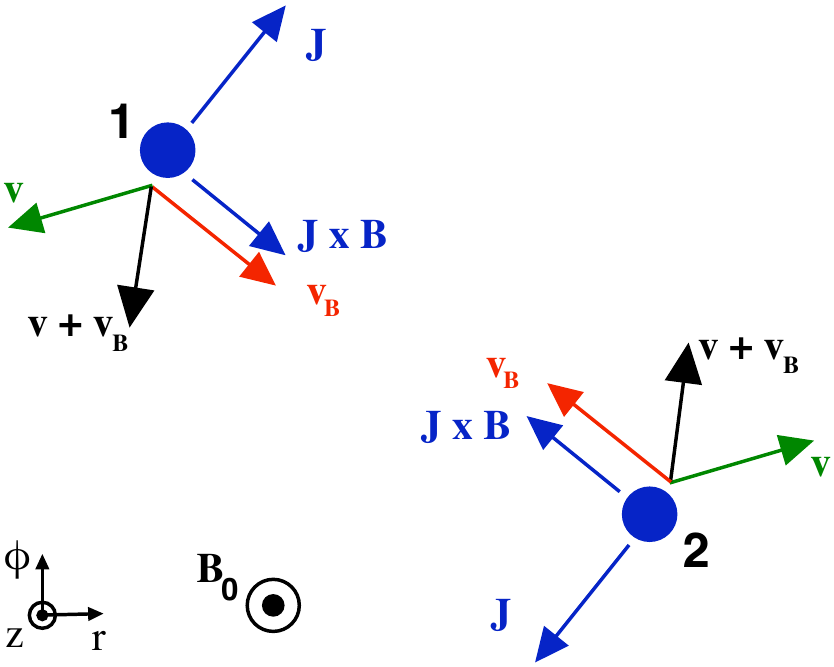}
    \caption{Plan view of the effect of ambipolar or Ohm diffusion
    on the development of the MRI. The current density and associated
    magnetic stress on the fluid at points 1 and 2 of Fig.
    \ref{fig:MRI-sketch} are indicated in blue.  Because of finite
    diffusivity the field lines drift through the fluid at velocity $\vv_\textrm{B}$ (\emph{red}
    vectors) in the direction parallel to $\J\cross\B$ (see eq
    \ref{eq:ion-mom}).  The net drift of the field line with respect to
    Keplerian rotation, $\vv+\vv_\textrm{B}$ is indicated by the \emph{black}
    vector.  In this case the net drift is marginally inwards
    at point 1. At
    point 2, the current density and magnetic stresses are reversed, so
    the field drifts in the opposite direction.  As a result the field
    lines drift inwards and outwards at reduced rates at points 1 and
    2 respectively, and the instability proceeds more slowly than in
    ideal MHD.}
    \label{fig:MRI-drift-AD}
\end{figure}

Fig.\ \ref{fig:MRI-drift-AD} shows the field line drift (red vectors) for the fluid configuration depicted in Fig.\ \ref{fig:MRI-sketch}, but now incorporating the effect of ambipolar or Ohm diffusion.  The black vectors indicate the total drift of the magnetic field in the local Keplerian frame, and correspond to the vector sum $\vv+\vv_{\rm B}$ (see eq \ref{eq:vB}). 
In this case the net effect of the drift is to reduce the radial and
azimuthal stretching of the field, and therefore -- as one might guess -- reduce the degree of instability.

In the Hall limit, collisions with the neutral gas are frequent and strong
enough to decouple the ions from the field, but the electrons -- which have
a smaller collision cross section with the neutrals and a higher
charge-to-mass ratio -- remain well coupled.  The magnetic field is then
frozen into the electrons, and these drift together through the neutrals
and ions, which remain tightly-coupled by collisions.  In this limit, the
field drift speed through the neutrals is given by the ion-electron drift
and hence is antiparallel to the current density,
\begin{equation}
    \vv_{\rm H} = \vv_e - \vv_i = - \frac{\J}{e n_e} \,.
    \label{eq:vB-hall}
\end{equation}
In eq.  (\ref{eq:vB-hall}), $n_e$ ($= n_i$) is the electron number
density and $e$ is the elementary electric charge.  

\begin{figure}
    \centering
     \includegraphics[scale=0.8,trim= 0 0 0 0]{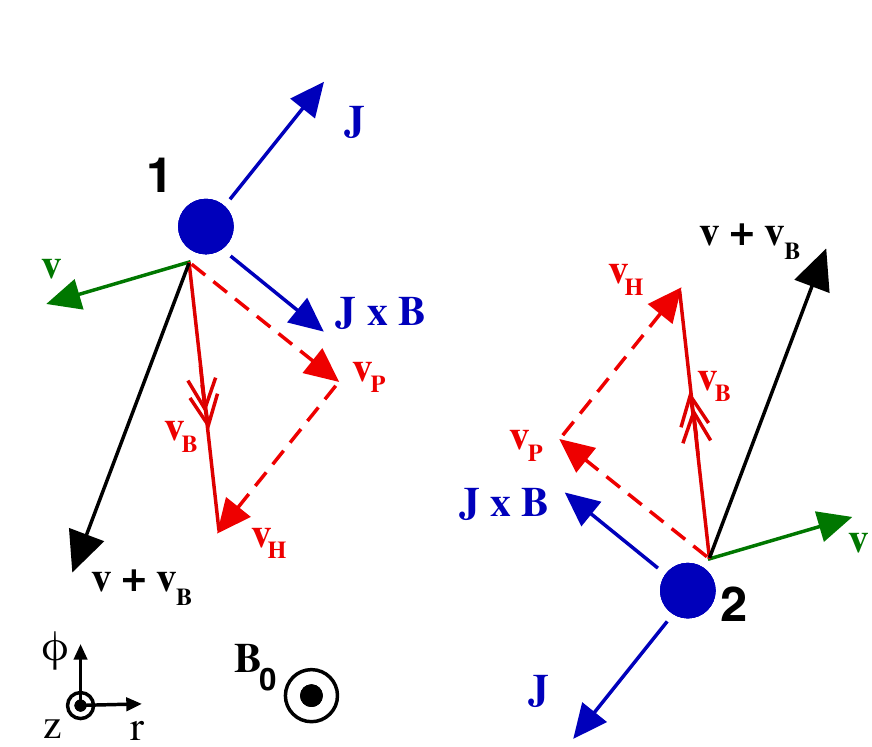}
   \caption{Plan view of the effect of magnetic diffusion 
	[incorporating the ambipolar, Ohm and Hall ($> 0$) terms] on the
	development of the MRI. The current density and associated magnetic
	stress on the fluid at points 1 and 2 of Fig.  \ref{fig:MRI-sketch} are
	indicated in blue.  At point 1 the magnetic tension provides some
	radial support against gravity and supplies a negative torque.  As a
	result, the material here is sub-Keplerian and moving inwards, as
	indicated by the green vector (which denotes departure from Keplerian
	rotation).  Because of 
	the
	finite diffusivity, the field lines drift
	through the fluid (\emph{red} vectors) with a velocity determined by
	the magnitudes of the ambipolar, Ohm and Hall diffusivities (see eq
	\ref{eq:vB}).  Ambipolar and/or Ohm diffusion contributes a drift
	$\vv_{\rm P}$ parallel to $\J\cross\B$, while the Hall drift $\vv_{\rm
	H}$ is antiparallel to the current density $\J$.  The net drift of the
	field line with respect to the Keplerian rotation, $\vv+\vv_{\rm B}$,
	is indicated by the \emph{black} vector.  In this case the net drift is
	inwards.  At point 2, the current density and magnetic stresses are
	reversed, so the field drifts in the opposite direction.  As a result
	the field lines drift inwards at point 1 and outwards at point 2 and the
	instability proceeds.}
    \label{fig:MRI-drift-a}
\end{figure}

More generally, when all diffusion mechanisms are taken into account the drift
velocity $\vv_{\rm B}$ of the field lines through the fluid is the sum of
these two orthogonal contributions (see eq \ref{eq:vB} below).  The
implications for the \mri\ are sketched in Fig.\
\ref{fig:MRI-drift-a}, which now includes a Hall drift antiparallel to
$\J$.
The generic geometry of the MRI means that the Hall drift at point 1
is directed inwards and retrograde in azimuth, and it is outwards and
prograde at point 2.  The net effect is to exacerbate the radial
buckling of the field, but to reduce the buckling in the azimuthal
direction.  We show in the next section that it is the radial effect
that is critical, as azimuthal field is always created out of the
underlying  Keplerian shear of the disc.  So Hall diffusion in this
case 
(e.g. when the magnetic field is aligned with the angular velocity vector of the disc)
tends to be destabilising.

\begin{figure}
    \centering
     \includegraphics[scale=0.8,trim= 0 0 0 0]{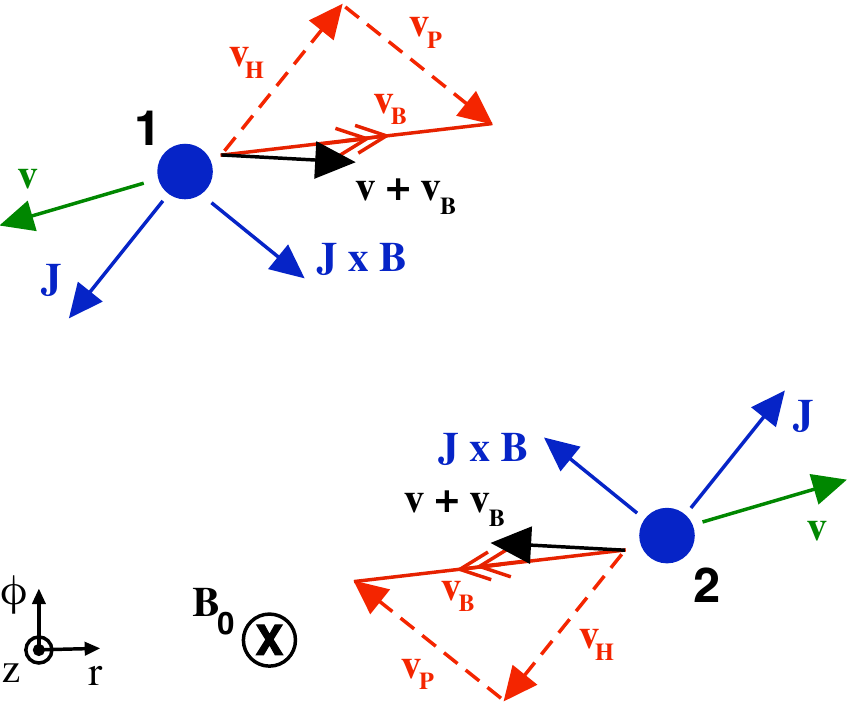}
    \caption{As for Fig.\ \ref{fig:MRI-drift-a}, but after a global
    reversal of the magnetic field, as would be expected if the
    initial magnetic field was antiparallel to the disc's angular
    velocity vector.  In this case, the magnetic stresses are unaffected --
    and so, therefore, are the fluid velocity and Ohm/ambipolar
    drift current density.  However, the current density is reversed,
    and therefore so is the Hall drift.  The magnetic field now drifts
    outwards at 1 and inwards at 2, reducing the radial buckling.  The MRI is
    partly or entirely suppressed in these circumstances (see text).}
    \label{fig:MRI-drift-b}
\end{figure}

This is not always the case, however, because the direction of the current
density, and therefore of the Hall diffusion, reverses upon global reversal of
the magnetic field.  This situation is illustrated in Fig.\
\ref{fig:MRI-drift-b}.
The magnetic stresses, the fluid velocity, and the field-line drift
associated with Ohm and ambipolar diffusion are not affected by this
reversal.
Note, however, that Hall diffusion now acts to stabilise the disc by acting against
the radial buckling of the field that would otherwise be driving the fluid
motions.

To summarise, Ohm and ambipolar diffusion are stabilising effects,whereas Hall diffusion may be destabilising or stabilising depending on whether the initial vertical magnetic field is parallel or antiparallel to the rotation axis, respectively.  This asymmetry reflects the fundamental asymmetry in the microscopic properties of the positive and negative charged species in the fluid in this limit.

\section{Linear Analysis of the MRI}
\label{sec:formulation}
We consider a small region of an axisymmetric, geometrically thin and nearly Keplerian disc, threaded by a vertical magnetic field, with sound and Alfv\'en speeds ($c_{\rm s}$ and $v_{\rm A}$) at the mid-plane that are both small compared to the local Keplerian speed $v_{\rm K}$. We assume that radial gradients are on the scale of $r$ and neglect vertical stratification of the initial equilibrium state, so that our analysis only holds near the mid-plane at heights $z\ll c_s /\Omega$. The initial state is in Keplerian rotation with a uniform density, pressure, and vertical magnetic field $\B=sB\zh$ (where $s=\pm1$). We linearise the equations around this state and seek solutions for axisymmetric perturbations of the form $\exp(\nu t - ikz)$. The equations for perturbations in density, pressure and $v_z$ form a separate system that describes vertically-propagating sound waves. The system of linear equations in the remaining perturbations involve fluctuations in the $r$ and $\phi$ components of $\B$, $\vv$ and $\vv_{\rm B}$.

A derivation of the dispersion relation is outlined in Appendices 
\ref{app:formulation} and \ref{app:analysis}.  This dispersion relation was first derived by \cite{W99}, and later extended to more general geometries in the Hall-Ohm limit \citep{BT01}, and then including ambipolar diffusion \citep{D04};  the ambipolar diffusion limit was also considered by \citet{KB04}.   Our emphasis here is on the dependence of the instability on the Pedersen and Hall diffusivities $\etaP$ and $\etaH$; a detailed analysis is presented in the appendices.  An overview is provided by Fig.\ \ref{fig:locii}, which illustrates the qualitative changes in the growth rate vs wave number curves for different choices of the diffusivities.  

First, note that the ideal-MHD limit holds at the origin (i.~e.~$\etaH=\etaP=0)$, the Hall MHD limit holds along the horizontal axis ($\etaP=0$) and the Ohm or ambipolar diffusion limits hold along the vertical line ($\etaH=0$).  Recall also that only the half plane $\etaP \geq 0$ is physically relevant.  It turns out that there are three distinct forms of the resulting $\nu(k)$ curve, corresponding to regions labelled I, II and III in the plane, as illustrated in Fig.\ \ref{fig:locii}.  In region I wave numbers less than a cutoff $k_c$ are unstable with the maximum growth rate attained at an intermediate wave number $k_0$ (see eqs \ref{eq:kc}, \ref{eq:nu0-1}, and \ref{eq:k0}).  In region II all wave numbers are unstable with the maximum growth rate still occurring at finite wave number. Finally, in region III  all wave numbers are unstable and the maximum growth rate is approached asymptotically as $k\rightarrow\infty$.
\begin{figure*}
	\centering
	\includegraphics[scale=0.8,trim= 30 0 0 0]{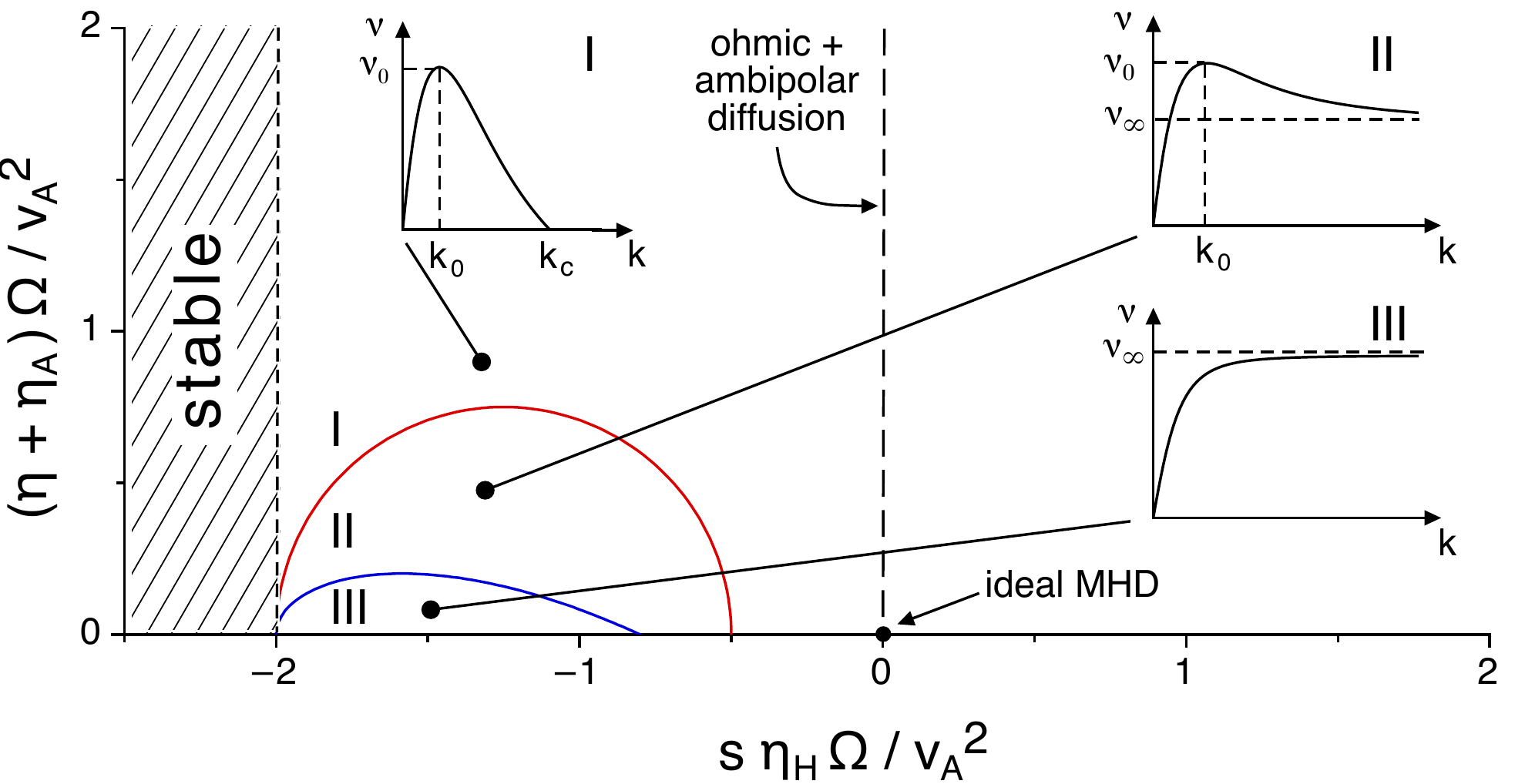}
	\caption{Schematic dependence of the behaviour of the MRI on the Hall and  Pedersen (Ohm+ambipolar) diffusivities, $\etaH$ and $\etaP$, for an initial magnetic field $sB\zh$ in a Keplerian disc, where $s=\pm1$ is the sign of $B_z$.  The field is assumed to be weak so that stratification can be neglected.  Physical values of the  Pedersen conductivity $\etaP$ are non-negative, and there are no unstable modes for Hall diffusivities $s\etaH\Omega/v_A^2\leq -2$.  The unstable region $s\etaH\Omega/v_A^2 > -2$ is subdivided according to the dependence of the growth rate on wave number $k$, which is sketched in the insets labelled I--III.  In region I (outside the red locus given by eq.\ \ref{eq:kcinf-locus}), wave numbers less than a cutoff $k_c$ are unstable with the maximum growth rate $\nu_0$ attained at $k_0$ (see eqs \ref{eq:kc}, \ref{eq:nu0-1}, and \ref{eq:k0}).  Between the red and blue loci (region II) all wave numbers are unstable with the maximum growth rate still occurring at finite wave number. Within the blue locus given by eq.\ \ref{eq:k0inf-locus} (region III)  all wave numbers are unstable and the maximum growth rate is approached asymptotically as $k\rightarrow\infty$.} 
	\label{fig:locii}
\end{figure*}
\begin{figure*}
    \centering
    \includegraphics[scale=0.8]{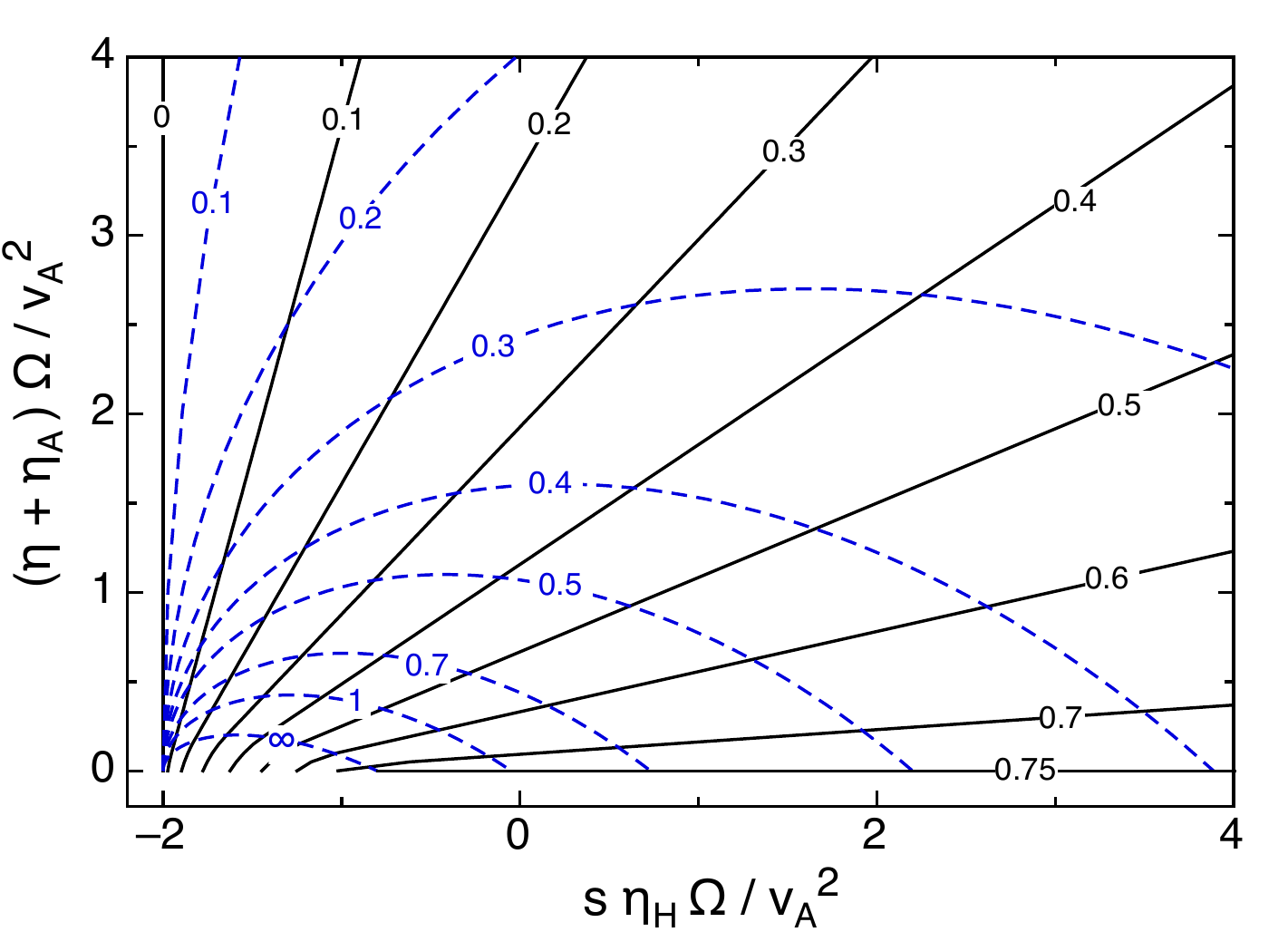}
	\caption{Contours of the maximum growth rate of the MRI (solid black lines, units 
$\Omega$) and corresponding	wave number (dashed blue lines, units $\Omega/v_A$) as a function of Hall and  Pedersen (Ohm+ambipolar) diffusivities.   The innermost blue contour 
shows where $k_0$ becomes infinite, corresponding to the boundary between regions II and III in Fig.\ \ref{fig:locii}.  Within this region the fastest growth rate is approached asymptotically as $k\rightarrow\infty$; this transition is responsible for the curvature of the black contours within this region (see eqs \ref{eq:nu0-1} and \ref{eq:nu0-2}).}
    \label{fig:contours}
\end{figure*}

Having delineated these three regions, we now consider how the critical
wave number $k_c$, fastest growth rate $\nu_0$ and corresponding wave
number $k_0$ vary across the entire $\etaP$--$\etaH$ plane.
Contours of the growth rate and wave number of the most unstable mode are plotted in this plane in Fig.\ \ref{fig:contours}.    
The growth rate increases clockwise, from $0\,\Omega$ along the vertical line $s\etaH=-2$ up to $0.75\,\Omega$ for the horizontal line $\etaP =0$ for $s\etaH>-4/5$.  In the absence of Hall diffusion, the maximum growth rate $\nu_0$ declines with increasing (Ohm and/or ambipolar) diffusivity (e.g. moving vertically upwards through the $s\etaH = 0$ point in the horizontal axis), with $\nu_0 \approx\textstyle{\frac{3}{4}\etaP^{-1}}$ for $\etaP\gg 1$.  The most important effect of Hall diffusion, apparent from Fig.\ref{fig:contours}, is that \emph{the growth rate of the MRI exceeds $0.3\,\Omega$ for $s\etaH\ga \etaP$, even for arbitrarily large $\etaP$}.  More generally, the addition of Hall diffusion at fixed $\etaP$ increases the growth rate if $s\etaH>0$ and decreases it when $s\etaH<0$.  For large values of $\etaP$, eq (\ref{eq:nu0-1}) shows that $s\etaH/\etaP \approx 24\nu_0/(9-16\nu_0^2)$.  It is this fact that has the potential to modify the extent of dead zones in protoplanetary discs, as we explore in \S \ref{sec:ppds}.

The wave number of the fastest growing mode (blue contours in Fig.\ \ref{fig:contours}) decreases as the diffusivity is increased. Again, the contours are not arranged so that the highest wave numbers occur in the ideal-MHD limit, but to the 
$s\etaH < 0$ side, within the boundary between regions II and III (traced by the $k_0=\infty$ contour).

\begin{figure}
    \centering
    \includegraphics[width=8.5cm]{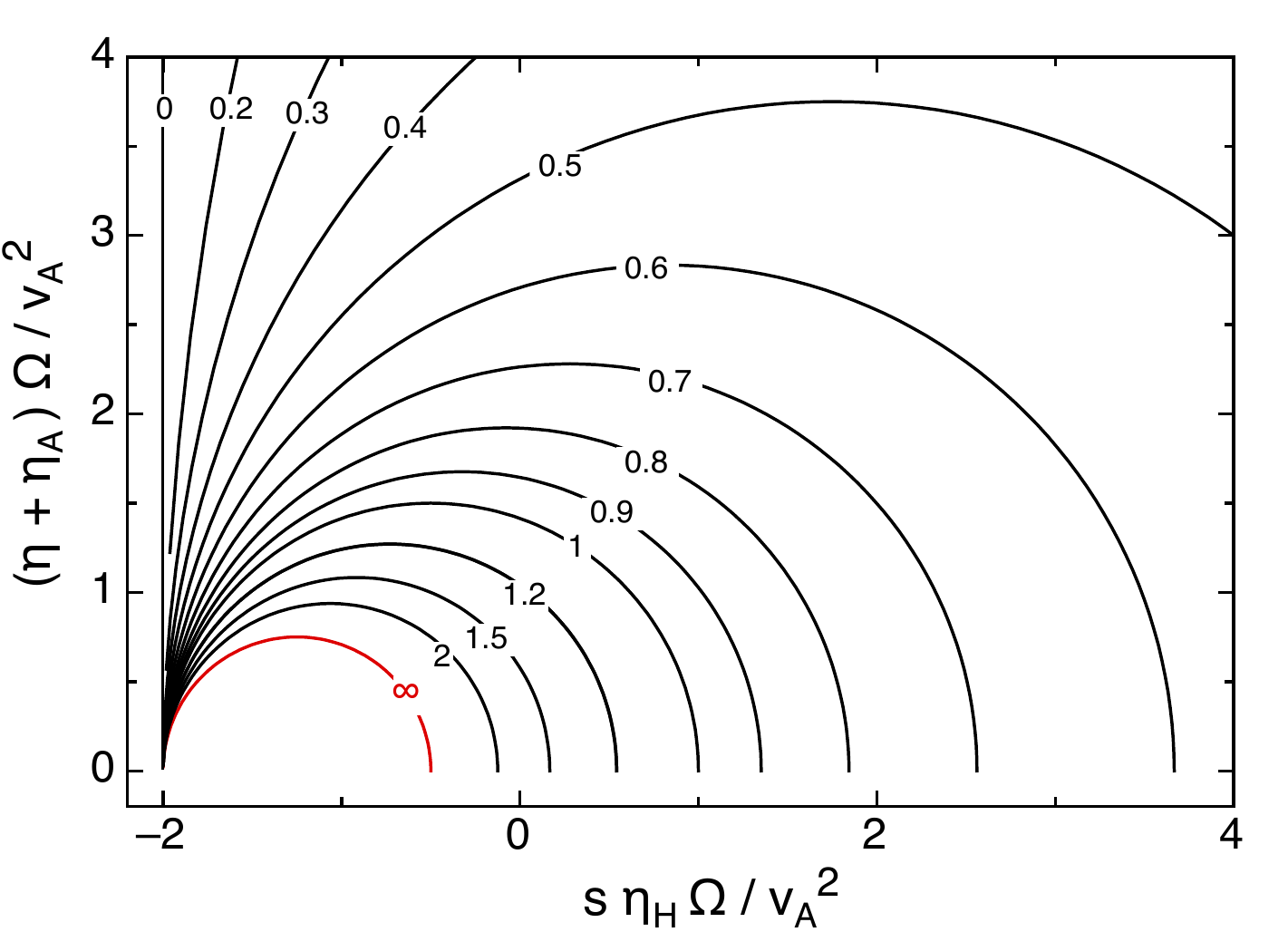}
    \caption{Contours of the critical wave wave number $k_c $ below
    which instability sets in (units $\Omega/v_A$).  The red innermost contour shows where
    $k_c$ becomes infinite and all wave numbers are unstable,
    corresponding to the boundary between regions I and II in Fig.\
    \ref{fig:locii}. }
    \label{fig:kcrit}
\end{figure}

 Contours of constant $k_c$ 
are semicircles, as plotted in  Fig.\ \ref{fig:kcrit}.
While the range of unstable wave numbers is reduced for large values of  
$s\etaH$ and $\etaP$, as one might expect, the range is not maximised in the
ideal limit (ie.  at the origin) but in regions II and III, bounded by the
$k_c=\infty$ contour.

In general we note that increasing $\etaP$ decreases the maximum growth rate and the characteristic wave numbers, whereas increasing $s\etaH$ above $-2$ increases the maximum growth rate and may either increase (when $s\etaH+1\la\etaP$) or decrease (when $s\etaH+1\ga\etaP$) the corresponding wave number.

Overall, these patterns place the ideal, Ohm (or ambipolar) and Hall regimes in context, and for the first time we see an overview of the effect of magnetic diffusivity on the linear MRI. In particular, there is nothing special about the Ohm/ambipolar limit, e.g.\ the behaviour of the instability in the presence of diffusion is not qualitatively different for $s\etaH=2$ vs $\etaH=0$. Even the ideal-MHD limit does not stand apart as remarkable, although it still holds a special place conceptually because flux freezing holds and it is easier to visualise. 

What does stand out is the part of the plane in the lower left, regions II and III, characterised by high wave numbers and the spraying out of the growth contours.  In this part of the plane, the instability operates in the ``cyclotron limit'' $\eta_\perp \sim 1$ and $k^2\gg 1$  (see appendix \ref{app:limits}), in which the instability arises in the competition between magnetic diffusion and advection of the field by the fluid; generation of $B_\phi$ from $B_r$ by the Keplerian shear flow is negligible. 
The
lack of any $k$ dependence in this regime occurs because both the magnetic diffusion and the magnetic stresses on the fluid (which are responsible for the fluid displacement) scale as $k^2$.
This short-wavelength, low-frequency limit corresponds to the cyclotron mode of the magnetised fluid, which has frequency $\omega_H = v_A^2/|\etaH|$ \citep{WN99,PW08}. This mode couples effectively to the Keplerian rotation as long as the sense of circular polarisation matches that of epicyclic motion, i.e. as long as $B_z\etaH < 0$. The other short-wavelength mode, the high-frequency whistler ($\omega \approx k^2 v_A^2/\omega_H$), is unable to couple effectively to the rotation \citep{W99}. 

\begin{figure}
    \centering
    \includegraphics[width=8cm]{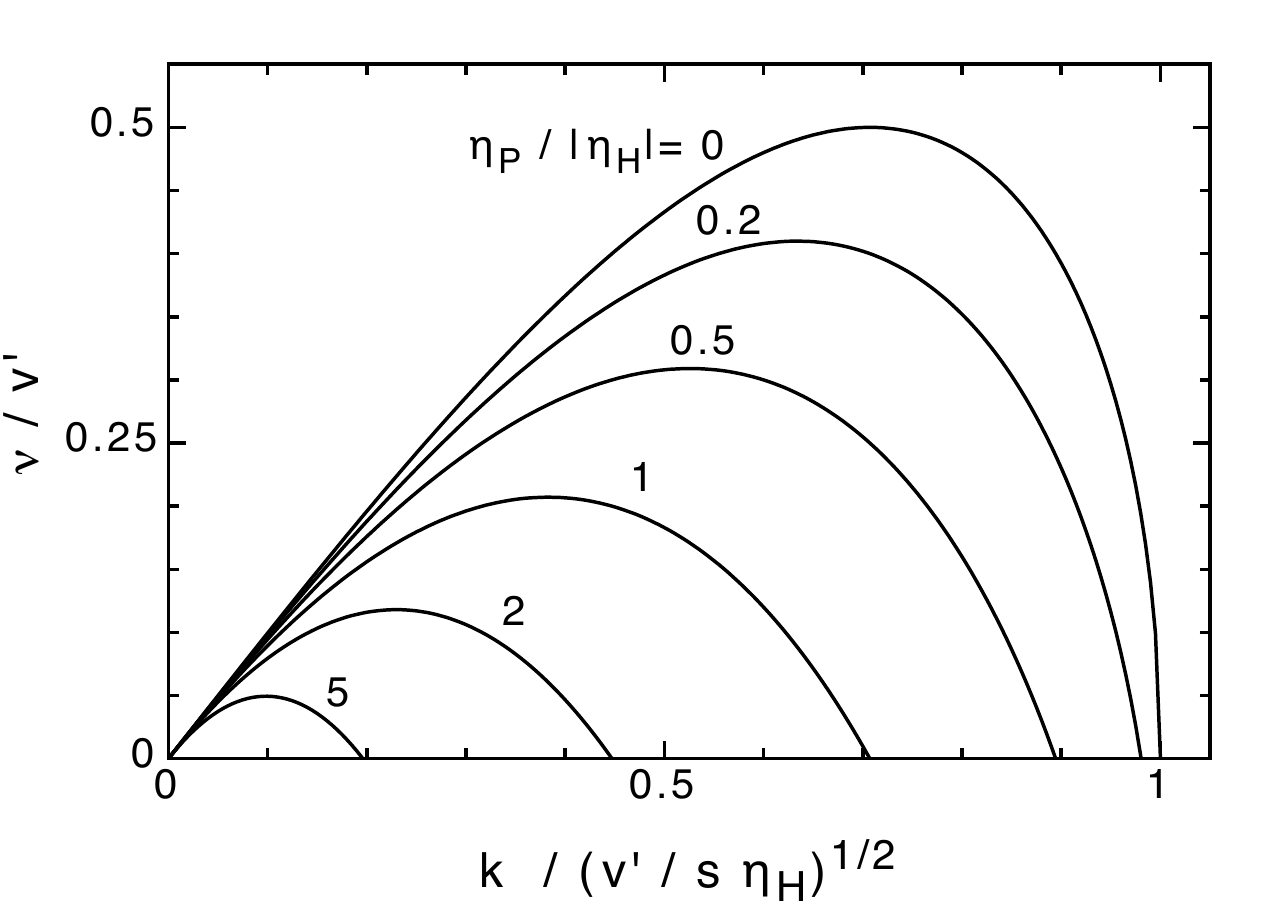}
    \caption{Growth rate versus wave number for the Hall-driven diffusive 
    instability for a mutually orthogonal magnetic field, shear velocity, and velocity 
    gradient $v'$. For Keplerian rotation and cylindrical geometry, the effective 
    velocity gradient is $v' = \thalf \Omega$. The curves correspond to 
    different values of the ratio 
    of the  Pedersen and Hall diffusivities.}
    \label{fig:nu-k-diffusion}
\end{figure}

The maximum growth rate contours emerging from regions II and III continue on to attain another limit when $\etaP^2+\etaH^2 \gg 1$, in which the field evolves in response to shear and diffusion, without significant feedback from the perturbations that it induces in the fluid flow.  Instability in this case relies on the keplerian shear flow generating $B_\phi$ from $B_r$, and the tendency of Hall diffusion to convert $B_\phi$ back into $B_r$.  This brings the potential destabilising effect of Hall diffusion in shear flows \citep{K08} to the fore and shows that it is quite independent of rotational effects -- i.e.~the Coriolis and centripetal acceleration -- that drive the MRI.  In Appendix \ref{app:limits} we obtain simple analytic expressions for growth rate in plane-parallel shear flows as a function of $k$ for arbitrary diffusivity, the results are plotted in Fig. \ref{fig:nu-k-diffusion}.
For more general field and wave vector configurations, ambipolar diffusion plays a similar role, albeit hindered by dissipation \citep[Pandey \& Wardle, in prep]{KB04,D04,K08}.

\section{Comparison with numerical simulations}
\label{sec:MRI-onset}

In this section we explore the relationship of our linear calculations to  the nonlinear, unstratified shearing box simulations of MRI-driven turbulence by  \citet{SS02a, SS02b}, which explicitly included both Ohm and Hall diffusion.  These authors' results could be interpreted as implying that the presence of Hall currents has little effect on the critical (minimum) degree of magnetic coupling  (the coupling between the neutral gas and the magnetic field) for the instability to operate\footnote{This field-matter coupling parameter is often referred to as  the magnetic Reynolds number $R_{\rm EM}$. As we expound below, however, it really corresponds to the Elsasser number $\Lambda$ (see discussion below and Table \ref  {table:numbers}).}. We show, however, that their calculations do not  yet probe deeply enough into the Hall regime for Hall diffusion to be  able to impact on the development and properties of the  instability.  In fact, their solutions are in agreement  with expectations from local and stratified linear analyses for the same values of the parameters describing the initial fluid conditions.  Finally, we use  the ratios of the different terms in the induction equation in order to  delineate the region of parameter space where Hall diffusion is expected to  substantially modify the growth rate, and spatial scale, of the MRI-unstable modes.  First, however, and for the sake of clarity, we discuss the notation used in the literature to characterise the fluid, and its magnetic activity, in diffusive environments.

\subsection{Fluid parameters}
\label{subsec:fluid}

\begin{table*}
\caption{Typical dimensionless numbers used to characterise the degree of coupling between the magnetic field and the neutral gas: Magnetic Reynolds No., Lundquist 
No., and Elsasser No. For comparison, we also show the Reynolds No.~(although this parameter combination does not involve the magnetic diffusivity and, therefore, 
does not measure the field-matter coupling). The fluid variables used to characterise the required scalings are the flow velocity ($v$), the Alfv\'en velocity ($v_
{\rm A}$), the kinematic viscosity ($\nu$) and the magnetic diffusivity $\eta$. As usual, $\Omega$ is the Keplerian angular frequency and $l$ is a typical, 
unspecified, length scale of the flow.}
\begin{tabular}{|c|l|l|c|c|c|c|}
\hline
Symbol & \multicolumn{1}{|c|}{Dimensionless} & \multicolumn{1}{|c|}{Ratio of} & \multicolumn{3}{|c|}{Scaling}  & Definition \\ 
& \multicolumn{1}{|c|}          {number}   & \multicolumn{1}{|c|}{physical quantities}     &  $V$ & $L$ &  $D$ & $VL/D$ \\
\hline\\
$R_{\rm e}$ & Reynolds & Inertial to viscous forces & $v$ & $l$ & $\nu$ & $vl/\nu$   \\[3pt]
$R_{\rm EM}$ & Magnetic Reynolds & Inertial to resistive time scales & $v$ & $l$ & $\eta$ & $vl/\eta$  \\[3pt]
$S$ &  Lundquist & Resistive diffusion to Alfv\'en time scales & $v_{\rm A}$ & $l$ &  $\eta$ & $v_{\rm A} l/\eta$  \\[3pt]
$\Lambda$ & Elsasser  & Lorentz to Coriolis forces & $v_{\rm A}$ &  $v_{\rm A}/\Omega$ & $\eta_\perp$ & $v_{\rm A}^2/\eta_\perp 
\Omega$  \\
\hline 
\end{tabular}
	\label{table:numbers}
\end{table*}

The growth and structure of the MRI are strongly dependent on the magnetic field strength, geometry, and the nature -- and magnitude -- of the magnetic diffusivity.  Different parametrizations have been used in the literature to characterise these properties, depending on the formulation of the problem and the adopted form of the induction equation.  Some authors use a tensor diffusivity, and the induction equation then takes the form 
\begin{eqnarray}
\delt{\B} &=& \curl \left (\vv \cross \B\right) - \curl \Big[
\eta \curl \B   
	\nonumber\\[6pt]
& & \quad \quad + \;\; \eta_{\rm H} (\curl \B) \cross \hat{\B} + \eta_{\rm A}
(\curl \B)_{\perp} \Big]
	\label{eq1:induction_drift}
\end{eqnarray}
(e.g.~\citealt{W07}).  Others prefer to use separate  equations for the charge carriers and the neutrals.  In the latter `multifluid' approach, when the ionised species are ions and electrons only (denoted by the subscripts `i' and `e', respectively), the induction  equation becomes (e.g. \citealt{K89,BT01,SS02a})
\begin{eqnarray}
\delt{\B} &=&  \curl \left (\vv \cross \B\right) -  \curl \Bigg[
\frac{c m_e \nu_{en} \J}{n_e e^2}
\nonumber\\[6pt]
& & \quad\quad +\;\; \frac{\J \cross \B}{e n_e} - \frac{(\J \cross \B) \cross \B}{c
\gamma_i \rho \rho_i} \Bigg] \,.
	\label{eq:induction_drift}
\end{eqnarray}
Comparison of equations (\ref{eq1:induction_drift}) and (\ref{eq:induction_drift}) yields the expressions given in (\ref{eq:eta}) to (\ref{eq:etaH}) for the  diffusivities. On the r.h.s.~of equation (\ref{eq1:induction_drift}), as well as in the second line of equation (\ref{eq:induction_drift}), the terms  (from left to right) denote the inductive (I), Ohm (O), Hall (H) and ambipolar diffusion (A) contributions to the evolution of the magnetic field, respectively. 

The following parameters have typically been used to characterise the magnetic properties of the fluid:
\begin{enumerate}
\item \emph{Field strength}.
This property is commonly measured either by the ratio of the Alfv\'en speed to the  isothermal sound speed ($v_{\rm A}/c_{\rm s}$), or by the  plasma beta parameter $\beta = (2/\gamma) c_{\rm s}^2/v_{\rm A}^2$,  where $\gamma$ is the adiabatic index of the fluid. 
\item \emph{Field-matter coupling}.
The degree of coupling between the neutral matter and the magnetic field  is, most generally, measured by a ratio of the type $V L/ D$ where $V$, $L$ and  $D$ represent characteristic speed, length and diffusion scales of the  flow, respectively.  In MRI studies, it is appropriate to take $V = v_{\rm A}$,  $L = v_{\rm A}/\Omega$ (the  characteristic wavelength of MRI-unstable modes in ideal-MHD conditions) and $D =  \eta_{\perp}$, the total perpendicular diffusivity [e.g. equation (\ref{eq:etaperp})]. The resulting ratio is the Elsasser number
\begin{equation}
\Lambda \equiv \frac{v_{\rm A}^2}{\Omega \eta_{\perp}} \,.
\label{eq:Lambda}
\end{equation}
This dimensionless number is formally defined as the ratio of the Lorentz force  ($\bmath{J} \times \bmath{B})/c$ to the Coriolis force $\propto \rho  (\bmath{v} \times \bmath{\Omega})$, which results in an expression of the  form $v_{\rm A}^2/(LV)\Omega$.  Adopting the magnetic diffusivity $\eta_\perp$ as  the typical magnitude of the LV factor in the denominator yields  expression (\ref{eq:Lambda}). 

The Elsasser number is widely used in geophysics, as in a geodynamo  process the magnetic field is thought to amplify until $\Lambda$ becomes of order unity (see, e.g.~\citealt{C10} and references therein).  This ratio was denoted $\chi$ in the  work of \citet{W99}, who  expressed it as
\begin{equation}
\chi \equiv \frac{B^2 \sigma_{\perp}}{\rho c^2 \Omega} \equiv
\frac{\omega_c}{\Omega} \,,
\label{eq:chi}
\end{equation}
where $\sigma_{\perp} = c^2/(4 \pi \eta_\perp)$ and $\omega_{\rm c}$ is the critical  frequency above which flux-freezing conditions break down, so  that the generic non-ideal MHD term in the induction equation dominates over the  inductive term.  When the field-matter coupling is characterised by the Elsasser number,  values $\gg 1$ ($\ll 1$) correspond to strong (weak) coupling.  Similarly, from equation (\ref{eq:chi}) it is clear that when $\Lambda \gg 1$, the magnetic field is strongly coupled to the  fluid at frequencies of order $\Omega$, the frequencies  of interest for the study of the MRI.

Note that, in general, the Elsasser number is different from the combination $v_{\rm A} l / \eta$, which compares the resistive diffusion time  $\tau_{\rm R} \propto l^2/\eta$ to the Alfv\'en time $\tau_{\rm A} \propto  l/v_{\rm A}$ (e.g. the Lundquist number) and from the ratio $v l/\eta$ (the  magnetic Reynolds number $R_{\rm EM}$, where $v$ is the fluid velocity). However, both  terms have been used in the literature to refer to $\Lambda$.  Table \ref{table:numbers} summarises the  definitions and typical scalings associated with the  dimensionless numbers discussed above, namely, the  magnetic Reynolds, Lundquist and Elsasser numbers.  For clarity we  also list the Reynolds number, although this  does not measure magnetic coupling, 
as it deals with the viscosity of the fluid instead of its magnetic diffusivity.

Finally, sometimes $c_{\rm s}$ is adopted as the characteristic velocity scale of the flow instead of $v_{\rm A}$, resulting in the ratio $c_{\rm s}^2/\eta \Omega$ (e.g.~\citealt{FSH00}).  

\item \emph{Diffusivity regime}.
This property characterises the importance of the different non-ideal MHD
terms in the induction equation. The ratios of each of the diffusive terms to the inductive term are typically used, which in a multifluid formulation are expressed as 
(e.g. \citealt{BT01}, \citealt{SS02a}):
\begin{equation}
\frac{O}{I} = \frac{\eta \Omega}{v_{\rm A}^2}  \equiv \tilde{\eta} \,, 
\label{eq:OI}
\end{equation}
\begin{equation}
\frac{H}{I} = \frac{X}{2} \,,
\label{eq:HI}
\end{equation}
and
\begin{equation}
\frac{A}{I} = \frac{\Omega}{\gamma_i \rho_i}  \,.
\end{equation}
Note that the inverse of the normalised Ohm resistivity $\tilde{\eta}^{-1}$ in equation (\ref{eq:OI}) has been referred to as the magnetic Reynolds number. However, as discussed 
above, this parameter combination is akin to the Elsasser number, with the difference that in the definition of equation (\ref{eq:Lambda}) we used the total 
perpendicular diffusivity $\eta_\perp$, instead of $\eta$. In equation (\ref{eq:HI}), 
\begin{equation}
X \equiv \frac{c B \Omega}{2 \pi e n_e v_{\rm A}^2}
\end{equation}
is the so-called ``Hall parameter", not to be confused with its
namesake, the ratio of the gyrofrequency to the collision frequency of ionized species $j$ (here either ions or electrons) with the neutrals, given by 
\begin{equation}
\beta_j \equiv \frac{eB}{m_jc}\frac{1}{\gamma_j \rho} \,.
\end{equation} 
Similarly, the relative importance of the 
non-ideal MHD terms can be measured by the following ratios (e.g. \citealt{BT01}, \citealt{SS02a})
\begin{equation}
\frac{H}{O} = \frac{eB}{m_e c} \frac{1}{\gamma_e \rho} \equiv \beta_e
= \frac{X}{2 \tilde{\eta}}     \,, 
\end{equation}
and
\begin{equation}
\frac{A}{H} = \frac{eB}{m_i c} \frac{1}{\gamma_i \rho} \equiv \beta_i \,.
\end{equation}

Using the tensor diffusivity notation, we can write
\begin{equation}
\frac{O+A}{I} = \frac{\etaP \Omega}{v_{\rm A}^2} \,,
\label{eq:etaOAI}
\end{equation}
\begin{equation}
\frac{H}{I} = \frac{\etaH \Omega}{v_{\rm A}^2} \,,
\label{eq:etaHI}
\end{equation}
and the relative magnitudes of the non-ideal MHD terms are, simply
\begin{equation}
\frac{H}{O} = \frac{\etaH}{\eta}
\label{eq:etaHO}
\end{equation}
and
\begin{equation}
\frac{A}{H} = \frac{\etaA}{\etaH} \,.
\label{eq:etaAH}
\end{equation}
From expressions (\ref{eq:etaOAI}) and (\ref{eq:etaHI}) it is clear that the Elsasser number is a measure of the ratio of the inductive term to the total non-ideal MHD terms in the induction equation. 
Furthermore,  for the vertical field geometry adopted here, the ambipolar diffusivity acts as a 
field-dependent resistivity\footnote{In other words, for this field geometry the current density $\J$ is perpendicular to $\B$ and the term $(\J \cross \B) \cross 
\B \equiv (\J \cdot \B) \B - B^2 \J = B^2 \J$ (see equation (\ref{eq:induction_drift})).} \citep[see ][]{BT01}. This property will be used in the next section to treat the Ohm and 
ambipolar diffusivities as an `Ohm-like' term when comparing the impact of the `Ohm-like' and Hall resistivities on the MRI.
\end{enumerate}

\subsection{Criterion for Hall diffusion to affect the MRI}

\begin{figure}
    \centering
    \includegraphics[scale=0.8]{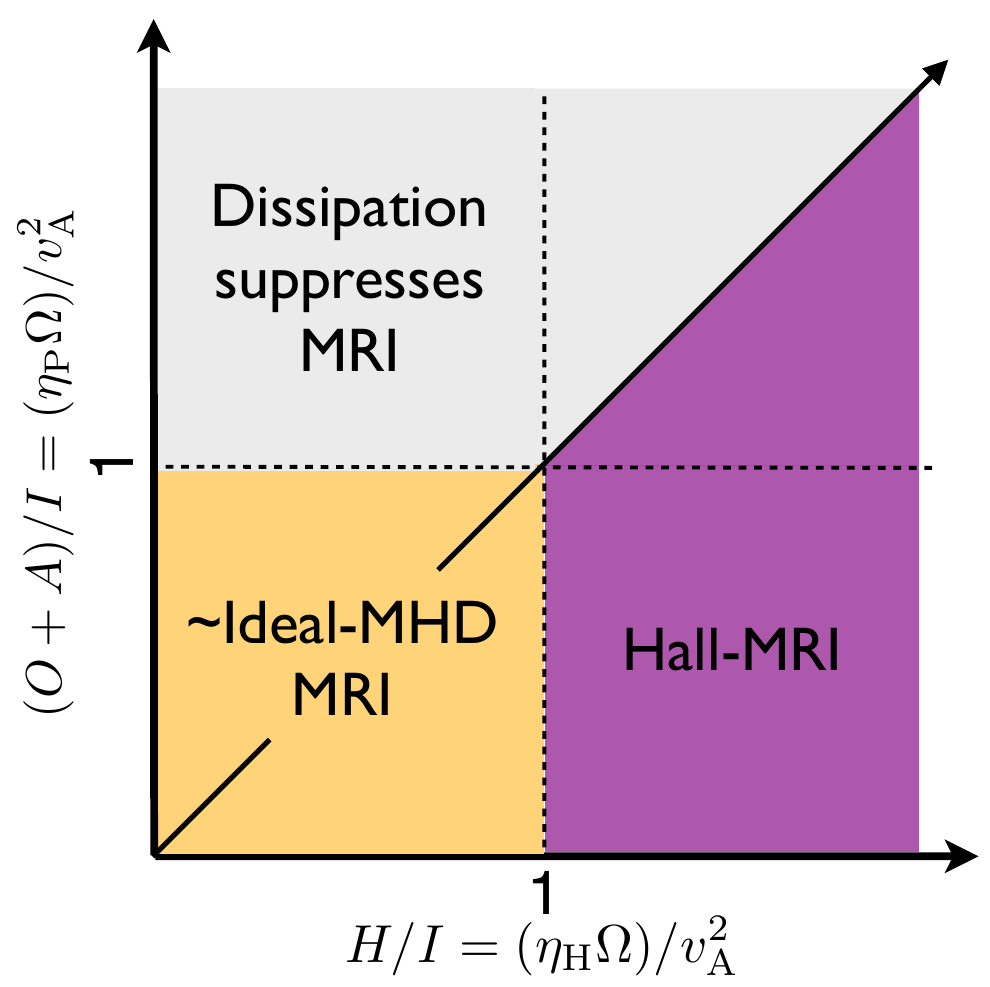}
    \caption{Regions of parameter space where the MRI is expected to grow, and dominant diffusion mechanism, in a (O+A)/I versus H/I (or, equivalently, the 
normalized  Pedersen versus Hall diffusivities) plane. Note that for the vertical field geometry considered here, the ambipolar and Ohm diffusivities can be combined 
into a generalized `Ohm-type' diffusivity (see text). In the lower-left (ochre) panel, magnetic diffusion is weak and the instability grows at a rate comparable to the ideal-MHD rate ($\sim 
\Omega$). Conversely, in the upper-left (grey) and lower-right (purple) panels, only one diffusivity term (Ohm-type and Hall, respectively) dominate over the 
inductive term in the induction equation. Ohm-type diffusivity suppresses the instability in the top-left quadrant (see footnote \ref{foot:bound}, however), but 
since Hall diffusion can also be destabilising (Section \ref{sec:diffusion}) the instability can still grow in the lower-right panel. Finally, both diffusivity 
components are important in the upper-right quadrant of the figure. In the top (grey) portion of this panel, Ohm-type diffusion overcomes Hall diffusion [H/(O+A) $< 
1$] and the MRI is expected to be suppressed. Hall diffusivity, however, is dominant [H/(O+A) $> 1$] in the bottom (purple) portion of the panel. In this region of 
parameter space, H/(O+A) \emph{and} H/I are both $> 1$ and the instability may proceed despite $\Lambda$ being less than unity. }
    \label{fig:hall_dom}
\end{figure}

\begin{figure}
    \centering
    \includegraphics[scale=1.0]{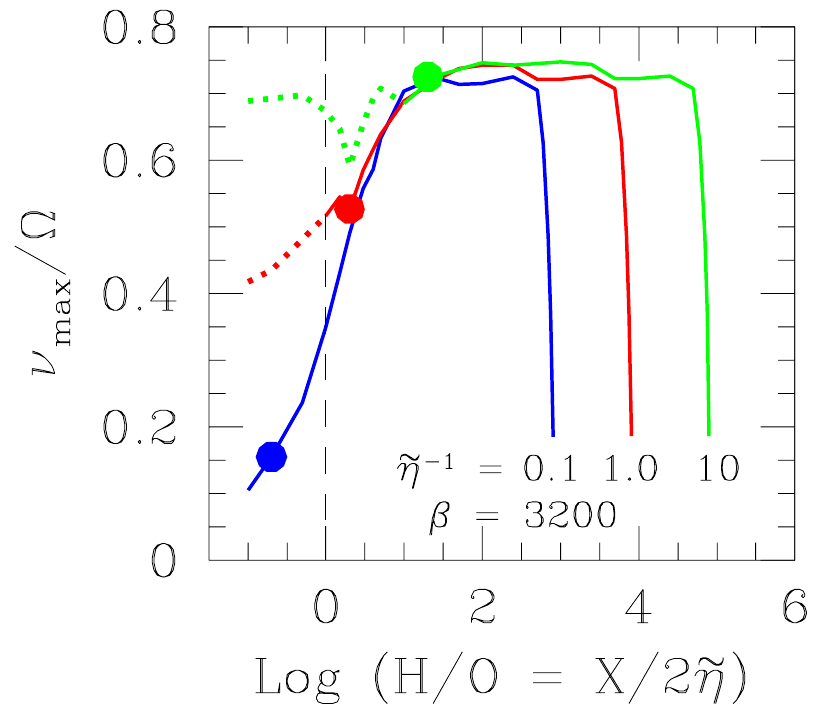}
    \caption{Growth rate of the fastest-growing MRI-unstable mode as a function of the ratio of the Hall to Ohm diffusion terms in the induction equation [H/O $= X/
(2 \tilde{\eta})$]. Calculations correspond to a stratified disc with $\beta \equiv (2/\gamma) c_{\rm s}^2/v_{\rm A}^2 = 3200$ (or $v_{\rm A}/c_{\rm s} = 0.02$). 
Each curve corresponds to a different value of $\tilde{\eta}^{-1}$ (= I/O): 
    0.1 (blue), 1.0 (red) and 10 (green).  A solid (dotted) line is used to plot solutions for which the ratio H/I $> 1$ ($< 1$).  Note that solutions for which Hall 
diffusion is dominant lie in the solid-line portion of the curves (H/I $> 1$) \emph{and} to the right of the vertical dashed line (H/O $> 1$). Superimposed (filled 
circles) are the linear growth rates corresponding to the solutions presented in Fig.~9 of \citet{SS02b} for $X = 4$. Note that the calculation for $\tilde{\eta}^
{-1} = 0.1$ and $X = 4$ satisfies H/O $= 0.2$ and is \emph{not} in the Hall-dominated region of parameter space. As a result, the MRI is significantly damped by Ohm 
diffusion. On the contrary, the instability can grow at about the ideal-MHD rate when H/O $> 1$ even when $\tilde{\eta}^{-1} = 0.1$.}
    \label{fig:Sano_comp}
\end{figure}

We now use the ratios given in the previous subsection to constrain the region
of parameter space where the MRI is expected to grow, as well as to
determine a criterion for Hall diffusion to substantially modify its properties.  First,
however, we discuss the results of the work of \citet{SS02b} on
this topic.  These authors characterised the Ohm and Hall
terms by their magnitudes relative to the inductive term in the
induction equation (e.g.~via eqs.~\ref{eq:OI} and \ref{eq:HI}).  Furthermore, they
considered that Hall diffusion was dominant when the ratio of the Hall to
inductive terms was larger than unity (H/I $> 1$).  Fig.~9 of \citet{SS02b}
shows the saturation level of the Maxwell stress (normalised by the initial
gas pressure), as a function of the initial (subscript `0') of the inverse Ohm resistivity ($\tilde{\eta_{\rm 0}}^{-1}$) -- called the magnetic Reynolds number -- 
for different values of the initial
plasma beta and Hall parameters ($\beta_{\rm 0}$ and $X_{\rm 0}$,
respectively).  

Their results indicate that when $\tilde{\eta_{\rm 0}}^{-1} \geq 1$,
the normalised saturated value of the Maxwell stress is of the order of
$0.1$ (with some scatter depending on the adopted $\beta_{\rm 0}$ and
$X_{\rm 0}$), and it is fairly independent of the actual value of
the resistivity.  The actual saturated magnitude of the stress is larger
in the models with $X_{\rm 0} = 4$ with respect to the ones with $X_{\rm 0}
= 0$ (e.g.~no Hall diffusivity) or $X_{\rm 0} = -2$ (negative Hall
diffusivity, when the magnetic field is counter-aligned with the disc
angular velocity vector).  On the contrary, when the initial inverse resistivity $\tilde{\eta_{\rm 0}}^{-1}$ is less than
unity, the saturation level of the Maxwell stress decreases by 1-2
orders of magnitude with respect to the $\tilde{\eta_{\rm 0}}^{-1} > 1$ case.  This trend of the Maxwell stress at saturation with
$\tilde{\eta_{\rm 0}}^{-1}$ seems to be unaffected by the presence (or magnitude) of
the Hall diffusivity.  These results have been interpreted to show that the
Hall diffusivity does not change the critical (maximum) $\tilde{\eta}$
required for the instability to grow ($\tilde{\eta}_{\rm crit} \sim 1$), but it
enhances the saturated level of the Maxwell stress by a factor of a few.

Note that the calculations of \citet{SS02b} probe the region of parameter
space where Ohm and Hall diffusivity terms dominate over the inductive term
(${\rm I}/{\rm O} = \tilde{\eta}_{\rm 0}^{-1} < 1$ and H/I $= X_{\rm 0}/2 > 1$).
However, $|X_{\rm 0}| \leq 4$ in all the presented calculations.  As a result,
for $\tilde{\eta}_{\rm 0} > 2$, the ratio H/O $= X_{\rm 0}/(2 \tilde{\eta}_{\rm 0})$ is $< 1$ and the
dominant diffusion mechanism is Ohm-type.  For the calculations with
$X_{\rm 0} = 4$ and $\tilde{\eta}_{\rm 0}^{-1} = 0.1$, in particular, the ratio ${\rm H}/{\rm O} = 0.2$, which implies that the Hall term is too weak
to overcome the damping effect introduced by Ohm diffusion.  It is not
surprising, therefore, that the large drop in the saturated value of the
Maxwell stress, with respect to that associated with the solutions
satisfying $\tilde{\eta}_{\rm 0}^{-1} > 1$, is not significantly modified by the
presence of Hall diffusion.  Naively, and as confirmed by the linear
analysis, the Hall effect should substantially modify the growth rate of
the \mri\ provided that Hall diffusion dominates over the inductive term (i.e.\
$|X_{\rm 0}|\ga 2$) \emph{and also dominates Ohm diffusion} (i.e.\ $X_{\rm 0}\ga
2\tilde{\eta}_{\rm 0}$).  According to these criteria, the Hall term would be strong enough to
modify the presented results if $X_{\rm 0} \geq 20$ (for the $\tilde{\eta}_{\rm 0}^{-1} =
0.1$ case), a value likely to be well beyond what has been computationally
feasible so far.

The above considerations are summarised in Fig.~\ref{fig:hall_dom}, which
shows the regions of parameter space where the instability is expected to
grow -- and, if so, the dominant diffusion mechanism -- in a [(O+A)/I - H/I]
or, equivalently, the normalised  Pedersen and Hall diffusivities, plane.  As discussed at the end of Section \ref{subsec:fluid}, we use a generalised `Ohm' term, which should be understood to mean `Ohm + ambipolar'. 

Diffusivity effects are weak in the lower-left (ochre) quadrant of the
figure, as the inductive term dominates over both the Ohm and Hall
diffusion terms in this region of parameter space.  The MRI is then
expected to grow at a rate not significantly reduced from the ideal-MHD
rate\footnote{Note, of course, that the boundaries between these regions
are not sharp as this simplified diagram might suggest, and the MRI is expected to grow at reduced rate when
$\etaP v_{\rm A}^2/\Omega \sim 1$ (e.g.~near the border between the lower and upper-left
panels of the figure).  This also applies along the diagonal line
separating the Hall and Ohm-dominated regions in the upper-right
panel.\label{foot:bound}}.  Conversely, in the upper-left (grey) and
lower-right (purple) panels, only one diffusivity term is dominant over the
inductive term (the Ohm and Hall term, respectively).  When Ohm diffusion
dominates, the instability's growth is damped, as this diffusion mechanism
is always stabilising (See the discussion in Section \ref{sec:diffusion} and Fig.~\ref{fig:MRI-drift-AD}).  However, the Hall term can be either stabilising or
destabilising (see Figs.~\ref{fig:MRI-drift-a} and \ref{fig:MRI-drift-b}), so when it dominates the instability can still potentially
grow .  Finally, the upper-right
quadrant of the figure contains the region of parameter space where both
the Ohm and Hall diffusivity terms dominate over the inductive term.
Although both diffusivity terms are important with respect to the ideal-MHD
term in this region, only in the lower (purple) portion of the panel Hall
diffusion is strong enough to potentially overcome the damping effect
introduced by Ohm diffusivity, as {\it only} here the ratios H/(O+A) and H/I
are both $> 1$.  Note that this region of parameter space was not probed by
\citet{SS02b}, as their solutions incorporating Hall diffusion lie in the
upper (grey) portion of the panel, where Hall diffusion is weak in
comparison to Ohm diffusion.  The potentially destabilising effect of a
sufficiently strong Hall diffusivity (the purple region of the upper-right
quadrant in Fig.~\ref{fig:hall_dom}), therefore, remains to be explored by
numerical simulations.

In order to test these assertions, we computed the growth rate of the most
unstable MRI mode in a stratified, geometrically thin and axisymmetric
accretion disc, using the method described in \citet{SW03}, to which we
refer the reader for additional details.  Our results are depicted in
Fig.~\ref{fig:Sano_comp} for $\beta = 3200$ (or $v_{\rm A}/c_{\rm s} =
0.02$), as a function of the ratio H/O ($= X/2\tilde{\eta}$).  Each curve
corresponds to a different value of the ratio I/O ($= \tilde{\eta}^{-1}$), as
follows: $ \tilde{\eta}^{-1}$ = 0.1 (blue), 1.0 (red), and 10 (green).  The range of
solutions shown with solid lines satisfy $X > 2$, so
that the Hall term dominates over the inductive term (H/I $> 1$).  The solutions depicted with a dotted line correspond to $X < 2$ (H/I $<
1$).  Note also that to the left of the vertical dashed line the ratio H/O
= $X/2\tilde{\eta} < 1$ and Ohm diffusion is the dominant non-ideal MHD term.

From the considerations in the previous paragraphs, it is clear that the
solutions that probe the Hall-dominated region of parameter space should
lie in the solid segment of each curve {\it and} to the right of the
vertical dashed line.  Note, in particular, that even for $\tilde{\eta}^{-1} = 0.1$, the instability
grows at essentially the ideal-MHD rate ($\nu_{\rm max}/\Omega \approx
0.75$) if the Hall term is strong enough (H/O $> 1$, or to the
right of the vertical line).  For comparison, we superimpose filled circles on the curves
showing the linear growth rate of the most unstable mode
corresponding to $X = 4$ and $\tilde{\eta}^{-1} = 0.1$, 1 and 10 (the set of
parameters tested by \citealt{SS02b}).  Note that the solution with $X = 4$
and $\tilde{\eta}^{-1} = 0.1$ lies to the \emph{left} of the H/O $= 1$ line, and Ohm
diffusion suppresses the growth of the instability.  We conclude, therefore, that the relative saturation levels of the MRI simulations presented in \citet{SS02b} are entirely consistent with expectations based on a linear analysis, and did not probe the regime $H>\{I,O,A\}$ in which the hall term strongly modifies the linear MRI.

\section{Application to protoplanetary discs}
\label{sec:ppds}

We now illustrate the implications of Hall diffusion for the extent of MRI-driven activity in protoplanetary discs by applying the linear analysis from 
appendix \ref{app:analysis} to the minimum mass solar nebula at 1\,au
  
While a linear analysis cannot hope to give a good sense of the properties of the nonlinear turbulence the MRI drives, numerical simulations in the Ohm limit have shown that it does predict when such turbulence exists, and appears to be an excellent predictor of the vertical extent of dead zones in stratified simulations of protoplanetary discs \citep{TS08}.  As noted in the previous section, the Hall-Ohm simulations of \cite{SS02a,SS02b} are also consistent with the expectations based on the linear dispersion relation.  Thus, we use the local dispersion relation to examine the role of the magnetic diffusivity in determining the extent of MRI-active regions in protoplanetary discs.

The tendency of the MRI to manifest on successively larger scales in the presence of increasing diffusivity is limited by the finite thickness of the disc: the wave numbers of interest are bounded from below by requiring $kh>1$ where $h = c_s / \Omega$ is the disc scale height.   Thus we adopt as a local criterion for growth of the MRI that the dispersion relation (\ref{eq:dispersion_relation}) yields modes with $\nu >0$ and $kh>1$. From this we infer that the diffusivities must lie within the semicircular locus obtained from eq (\ref{eq:disp-circle}) with $\nu=0$ and $k=h^{-1}$, i.e.
\begin{equation}
    \left(\frac{s\etaH\Omega}{v_A^2} + 
    \frac{5}{4}-\frac{3}{4}\frac{c_s^2}{v_A^2}\right)^{2} + 
    \left(\frac{\etaP\Omega}{v_A^2}\right)^2  <
    \;\frac{9}{16}\left(1+\frac{c_s^2}{v_A^2}\right)^2
    \,.
    \label{eq:disp-critical}
\end{equation}
where $s=sign(B_z)$.
This criterion applies to any near- Keplerian disc threaded by a vertical magnetic field.  Its apparent simplicity belies the fact that the diffusivities are complicated (but calculable) functions of magnetic field strength, gas density, temperature, and the abundances of charged species.    Note also that the ratio $c_s^2/v_A^2$ plausibly ranges between 0 and the equipartition value
$2$ at the disc  mid-plane  but may be significantly higher away from the  mid-plane if the field is anchored at lower heights within the disc.

\begin{figure}
    \centering
    \includegraphics[width=8.5cm]{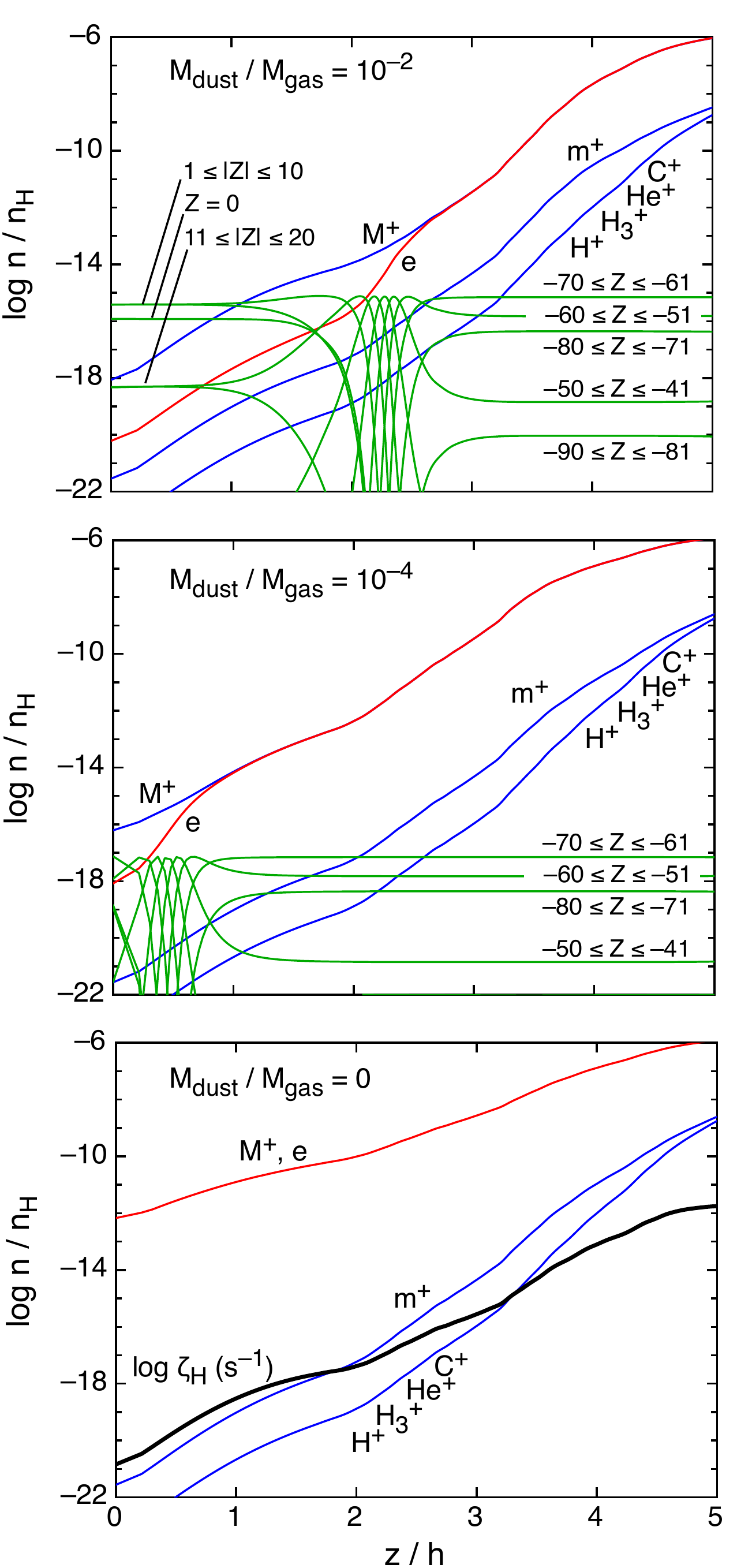}
     \caption{Fractional abundances of charged particles in the minimum mass solar nebula at 1\,au from the Sun as a function of height above the mid-plane, assuming that grains have radius 1\,\micron. The upper, centre, and lower panels correspond to dust to gas ratios $10^{-2}$, $10^{-4}$, and 0 by mass, respectively. The black curve in the lower panel shows the height-dependence of the ionisation rate per hydrogen nucleus due to stellar x-rays and interstellar cosmic rays (see text) assumed for all three models. The other curves give the fractional abundances of electrons (\emph{red}), light ions (H$^{+}$, H$_3^{+}$, He$^{+}$, C$^{+}$) representative molecular (m$^{+}$), and metal (M$^{+}$) ions (\emph{blue}), and grains (\emph{green}, labelled by charge state).}
    \label{fig:x1au}
\end{figure}

We specialise to protoplanetary discs by adopting the ionisation models of \citet{W07} for a standard minimum mass solar nebula disc ionised by cosmic-rays and stellar x-rays at 1\,au from the central star.  These models assume a column density of $1700\gpersc $
and temperature $T=280$\,K independent of height, with x-ray ionisation rate computed by \citet{IG99} and a standard interstellar cosmic-ray ionisation rate of $10^{-17}$\,s$^{-1}$\,H$^{-1}$ that is exponentially attenuated with depth as $\exp(-\Sigma / 96\gpersc )$.  We characterise the grain population by assuming a single 1\,\micron\ radius and varying the dust-to-gas mass ratio, crudely mimicking the effect of the settling of grains to the disc mid-plane.

The abundances of charged species for dust-to-gas mass ratios of $10^{-2}$, $10^{-4}$ and 0 are presented in Figure \ref{fig:x1au}.
The ionisation rate as a function of depth is plotted in the lower panel -- interstellar cosmic rays are the dominant source below 2 scale heights, above this ionisation by stellar x-rays dominates.  In the no-grain case, electrons and metal ions are the dominant charged species because the metal ions have the smallest recombination rate coefficient.  The top panel shows the effect of adding a population of 1\,\micron-radius grains with total mass 1\% of the gas mass: grains acquire a charge via sticking of electrons and ions from the gas phase.  Above $z/h\approx 2.5$ the grain charge is determined by the competitive rates of sticking of ions and electrons, with the Coulomb repulsion of electrons by negatively-charged grains offsetting their greater thermal velocity compared to ions \citep{Spit41, DS87}.  This leads to a gaussian grain charge distribution with mean charge (in units of $e$) 
$\langle Z_g \rangle  \approx -4akT/e^2 \approx-67$ and standard deviation 
$\approx \langle Z_g \rangle ^{1/2}\approx 8$.
  Most recombinations still occur in the gas phase.  The abundance of metal ions and electrons is reduced over the grain-free case for $z/h\la 5$ where the charge stored on grains becomes comparable to the electron abundance.  For $z/h\approx 2$--2.5, the abundances of ions and electrons have declined to the point that the majority of electrons stick to grain surfaces before they can recombine in the gas phase, and most neutralisations occur when ions stick to negatively charged grains \citep{NU80}.  Closer to the  mid-plane, the ionisation fraction is so low that most grains are lightly charged, so that Coulomb attraction or repulsion of ions and electrons by grains is negligible. Then ions and electrons stick to any grain that they encounter, with recombinations occurring on grain surfaces.  The middle panel of Fig.\ \ref{fig:x1au} shows what happens if 99\% of the grains are removed (e.g.~by settling to the  mid-plane).  The capacity of the grain population to soak up electrons from the gas phase is reduced a hundredfold, and the height below which grains substantially reduce the free electron density below the ion density moves downwards to the lowest scale height. 

\begin{figure}
    \centering
    \includegraphics[width=8.5cm]{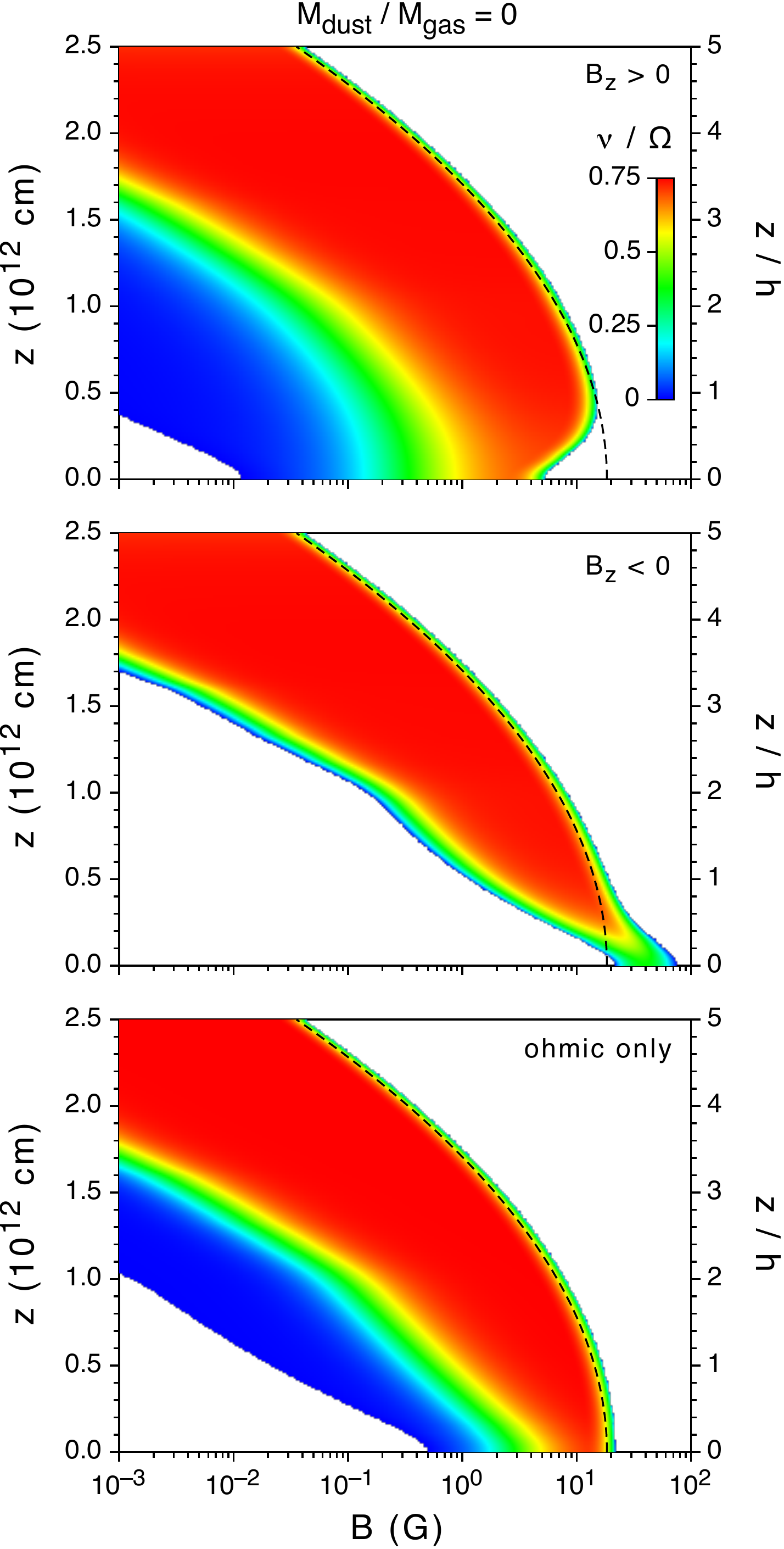}
    \caption{Colour shading shows the growth rate of the fastest growing local MRI mode at height $z$ above the  mid-plane in the minimum mass solar nebula at 1\,au as a function of the strength of an initially vertical magnetic field $B$. Dust grains are assumed to have settled to the  mid-plane  (see lower panel of Fig.\ \ref{fig:x1au}). The vertical wave number $k$ of the modes are required to satisfy $kh>1$ where $h$ is the disc scale height. The unshaded regions show stable combinations of height and field strength.    The top and middle panels correspond to the cases where the initial field and rotation axis are parallel ($B_z>0$) or antiparallel ($B_z<0$), respectively.  The lower panel shows the effect of artificially suppressing ambipolar and Hall diffusion, so including only Ohm diffusion.  In this case there is no dependence on the sign of $B_z$.  The dashed line in each panel indicates the value of the local equipartition magnetic field as a function of height above the  mid-plane.\label{fig:maxnu-f=0}}
\end{figure}

\begin{figure}
    \centering
    \includegraphics[width=8.5cm]{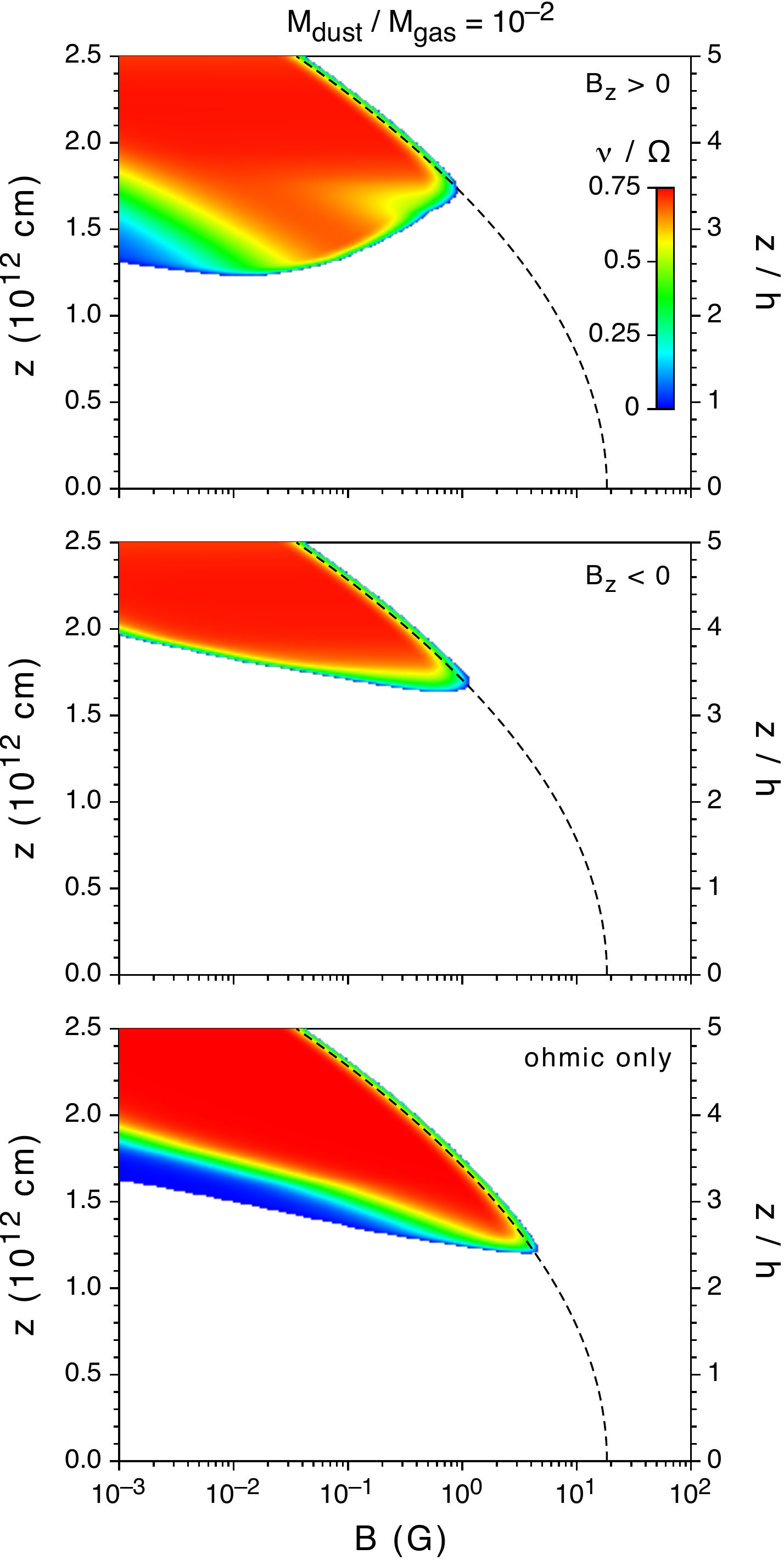}
    \caption{As for Fig.\ \ref{fig:maxnu-f=0}, but now including a 
    population of 1$\,\mu$m radius grains with total mass 1\% of the gas mass 
    (see upper panel of Fig.\ \ref{fig:x1au}).\label{fig:maxnu-f=1}}
\end{figure}

Unlike Ohm resistivity, the Hall and ambipolar diffusivities depend on the magnetic field strength as well as the charged particle abundances, so we consider field strengths ranging between $10^{-3}$--$10^2$\,G, encompassing the plausible range of values 
in the solar nebula (see \citealp{W07}).  For each choice of magnetic field we use the ionisation calculations to compute the diffusivities as a function of height.  Typically, ambipolar diffusion dominates at the surface and, for weak magnetic fields, Ohm diffusion dominates near the  mid-plane.  Between these regimes Hall diffusion dominates \citep{W07}.  With the diffusivity profiles in hand, we apply eq (\ref{eq:disp-critical}) to identify the fastest growing MRI mode that has $kh>1$ at a given height; the results for dust-to-gas mass ratios $0$ and $10^{-2}$ are displayed in Figs~\ref{fig:maxnu-f=0} and \ref{fig:maxnu-f=1} respectively.

First consider the case in which grains are absent (Fig.~\ref{fig:maxnu-f=0}).  The bottom panel shows the fastest growth rate (subject to 
$kh>1$) as a function of magnetic field strength and height if Hall and ambipolar diffusion are neglected and only Ohm diffusion is included.  In this case the departure from ideal MHD is measured by $\eta\Omega/v_A^2$ (the inverse of the Elsasser number; see section \ref{subsec:fluid}), which strongly decreases with height as the fractional ionisation and Alfv\'en speed both increase 
sharply away from the midplane.  For the ionisation profile plotted in Fig.\ \ref{fig:x1au} it turns out that $\eta\Omega/v_A^2$ is small near the surface (so that ideal MHD holds) and large near the  mid-plane (so that Ohm damping is severe).  As a result, near the surface the largest achievable growth rate is close to the ideal value 0.75\,$\Omega$, but declines rapidly below the height where $\eta\Omega/v_A^2 \sim 1$.  The height of the transition between these regimes declines with increasing field strength, simply because 
$\eta\Omega/v_A^2\propto B^{-2}$.  A second consideration is that the range of MRI-unstable wave numbers is bounded above by $k_c$ (see eq \ref{eq:kc}), and this must be larger than  $1/h$ if \emph{any} unstable modes are to exist with $kh>1$.  It is this criterion that provides the upper and lower envelopes to the unstable region in Figs.\ \ref{fig:maxnu-f=0} and \ref{fig:maxnu-f=1}.  Near the surface where 
$\eta\Omega/v_A^2 \ll 1$, $k_c$ is approximately $\sqrt{3}\Omega/v_A$, and as $v_A$ rapidly increases with height the unstable range shifts to wave numbers with $kh<1$ and it is no longer possible to find unstable modes that fit within a scale height.  This occurs when the magnetic pressure roughly exceeds the gas pressure.  By contrast, close to the  mid-plane where $\eta\Omega/v_A^2 \gg 1$, $k_c \approx \sqrt{3}v_A/\eta$ and the wave numbers are pushed out of the relevant range by the increasing diffusivity and the declining Alfv\'en speed as the  mid-plane is approached.

Next, we add the remaining magnetic diffusion terms, ie.\ ambipolar and Hall diffusion.  The latter depends on the sign of $B_z$, which we take to be positive or negative in the upper and middle panels of Fig.\ \ref{fig:maxnu-f=0} respectively.  Ambipolar diffusion dominates in the surface layers (see Fig.\ 5 of \citealt{W07}), and it acts just like Ohm diffusion when, as assumed here, the initial magnetic field is vertical
\footnote{This is not the case in more general geometries, e.g.~ambipolar diffusion may be destabilising when the field has vertical and toroidal components and the wave number has radial and vertical components (see \citealt{KB04} and \citealt{D04}).}, tending to damp the growth of the MRI and pushing the unstable 
modes to longer wavelengths.  Hall diffusion tends to dominate the lower layers, except for weak fields when Ohm diffusion dominates closer to the  mid-plane.  

For $B_z > 0$ Hall diffusion is destabilising and pushes the region where the fastest growth rate is close to 0.75\,$\Omega$ to greater depths.  The rapid decline to low growth rates (ie.\ from green to blue shading in Fig. \ref{fig:maxnu-f=0}) traces the transition from Hall-dominated to Ohm-dominated diffusion.  Below this, Hall diffusion acts to significantly extend the region of slow growth by modifying $k_c$, and this extends all the way to the  mid-plane for fields in excess of 10\,mG. 

On the other hand, when $B_z<0$ (ie.\ $\B$ is antiparallel to the rotation axis) Hall diffusion tends to stabilise the MRI while extending the unstable wavelengths to shorter wavelengths (see Fig.\ \ref{fig:contours}). As a result, super-equipartition fields are unstable near the  mid-plane.  We emphasise that \emph{irrespective of the magnitudes of the other diffusivities Hall diffusion is completely stabilising when $\etaH > 2 v_A^2/\Omega$.}This is responsible for the sharp cutoff at the lower boundary in the middle panel of Fig.\ \ref{fig:maxnu-f=0}.

When a full complement of 1\,\micron\ dust grains are present (i.e.~dust-to-gas mass ratio $10^{-2}$; charged species as in lower panel of Fig.\ \ref{fig:x1au}), the diffusivities are greatly increased near the  mid-plane because electrons are locked up by grains and rendered immobile.    Fig.~\ref{fig:maxnu-f=1} displays the same trends as in the zero-grain case, but the MRI-unstable region is now restricted to the upper layers of the disc, and in all cases the bulk of the disc is stable to the MRI.   While the differences between the three panels might appear less severe in this case, the strong density stratification means that there are orders of magnitude differences in the column density of the MRI-unstable region. 

\begin{figure}
    \centering
    \includegraphics[width=8.5cm]{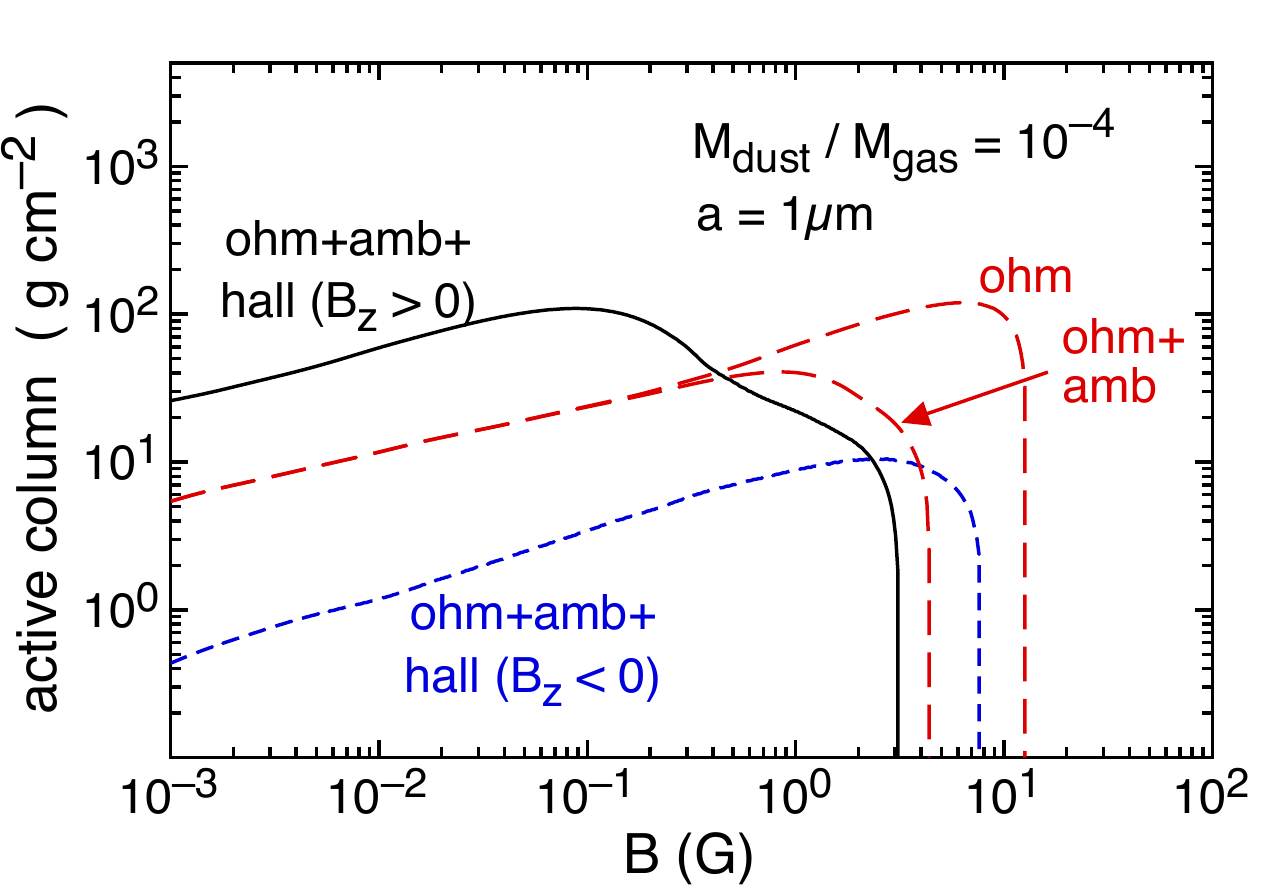}
    \caption{Column density of the MRI-active region at 1\,au in the minimum mass solar nebula as a function of magnetic field strength for different assumptions about magnetic diffusion (see text).  Grains are assumed to have radius 1\micron\ with 99\% having settled to the disc mid plane (i.e.~total mass only $10^{-4}$ relative to gas).  \emph{Solid black} and \emph{dashed blue} curves are for the magnetic field oriented parallel or antiparallel to the disc's rotation axis respectively.  \emph{Long-dashed red} curves indicate the effect of accounting only for Ohm or Ohm+ambipolar (i.e.~Pedersen) diffusion. \label{fig:active-sigma}}
\end{figure}

Fig.~\ref{fig:active-sigma} shows the magnetically-active column density as a function of field strength for different assumptions regarding the diffusivity when 99\% of grains are assumed to have settled (i.e.\  dust to gas ratio $10^{-4}$).
The active column density varies by 1-2 orders of magnitude depending on the accuracy of the treatment of magnetic diffusion.  First, neglect Hall diffusion and consider either Ohm diffusion alone or Ohm and ambipolar diffusion (i.e.~Pedersen diffusion) operating in concert (red long-dashed curves).  Ohm diffusion dominates ambipolar diffusion except for strong magnetic fields and low densities, so there is little difference between these two cases except for magnetic fields in excess of 1\,G where the additional damping due to ambipolar diffusivity suppresses the MRI.    Hall diffusion either increases or decreases the active column density by an order of magnitude depending on whether the initial magnetic field is pointing up or down; the maximum unstable field strength also varies considerably.

\begin{figure}
    \centering
    \includegraphics[width=8.5cm]{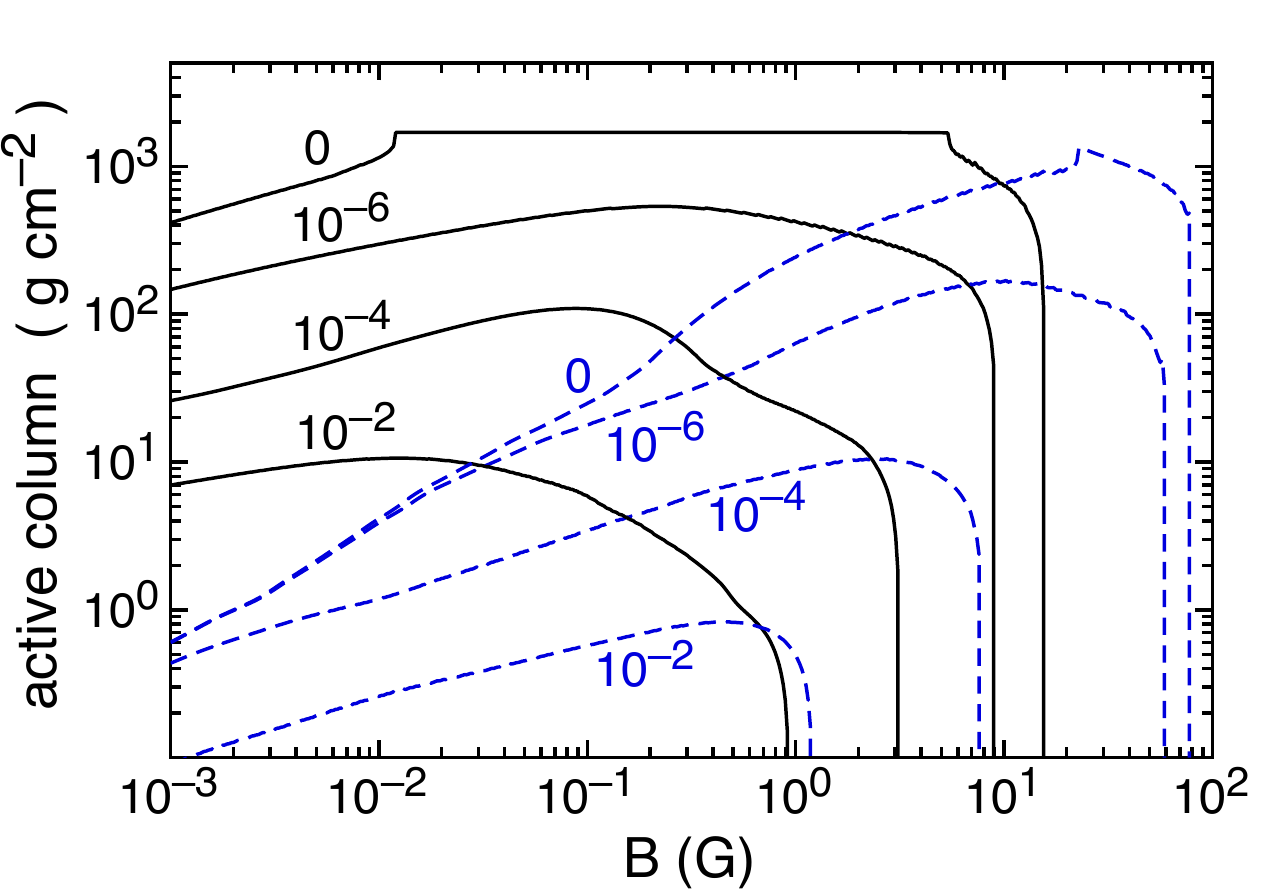}
    \caption{
    As for Fig. \ref{fig:active-sigma} but for dust-to-gas mass ratio varying from $0$ to $10^{-2}$.  Note that active column density is capped at the total column of the minimum mass solar nebula at 1\,au, i.e.~1700\,g\,cm$^{-2}$.\label{fig:active-sigma-dust}}
\end{figure}
Fig.~\ref{fig:active-sigma-dust} shows how the active column depends on grain abundance and magnetic field orientation.   In the absence of grains the entire disc cross-section is magnetically active for an upwardly-directed field in the range 0.01--5\,G.   On the other hand, when the field is pointing downwards, a minuscule fraction of the disc is active for weak fields, but the active column increases rapidly and encompasses the entire
cross section of the disc
for field strengths in the range 20--80\,G.   As one would expect, grains sharply reduce the active column density because of the reduction in mobile charge carriers. The continued extreme sensitivity to field alignment demonstrates that Hall diffusion still plays a critical role in determining the extent of magnetic activity regardless of the grain abundance.

The reduction in the active column when $B_z < 0$ occurs because Hall diffusion stabilises the disc against the MRI when $s\etaH<-2 v_A^2/\Omega$.
When the grains and ions are strongly coupled to the neutrals and the electrons are coupled to the magnetic field, the Hall diffusivity is given by eq (\ref{eq:etaH}) and we obtain a field-dependent criterion for stability on the fractional ionisation
\begin{equation}
    \frac{n_e}{n_H} \, \la \;\frac{1}{2} \, \frac{1.4 m_H c \, \Omega}{eB} \approx \frac{1.5\times 10^{-11}}{B(\mathrm{G})} \,\frac{(M/\mathrm{M_\odot})^{1/2}}{ (r/\mathrm{au})^{3/2}}\,.
    \label{eqn:xe-crit}
\end{equation}
This critical fractional ionisation is typically larger than the corresponding criterion for Ohm damping, so sets the lower boundary of the magnetically active region when $B_z < 0$.  
Note that once the grain abundance is low enough not to affect the ionisation fraction, the lower boundary reaches the midplane and the active column density saturates.

These results suggest that Ohm estimates of the column density of the magnetically-active layers in protoplanetary discs are in error by about an order of magnitude, systematically under- or overestimating the active column if the magnetic field is directed upwards or downwards, respectively.  

This conclusion rest on several assumptions.  Our neglect of stratification in the linear analysis is unlikely to be serious given that in the Ohm limit this approach yields excellent predictions of the extent of the depth of the turbulent layers \citep{TS08}. The restriction to a simple and somewhat degenerate geometry -- i.e~ to vertical fields and wave vectors -- is perhaps more suspect.  However, when ambipolar diffusion is unimportant,  a toroidal component of the magnetic field and a radial component of the wave vector can easily be accommodated: a rescaling of the full dispersion relation yields the identical dispersion relation (Pandey \& Wardle, in preparation).   The situation is less clear when ambipolar diffusion is involved, as in this case it stabilises or destabilises the MRI 
  \citep{KB04,D04}
  in an analogous manner to the Hall contribution for the vertical field considered here (Pandey \& Wardle, in preparation).  However, at $\sim1$\,au, ambipolar diffusion only dominates near the disc surface where the ionisation fraction is so high that diffusion is small in any case.   Finally, although the grain model is crude, the key property of the grain population is its capacity to soak up electrons, which is proportional to $\int a\,n(a)\,da$ \citep{W07}.  
  This can be used to scale our results to any grain size distribution.  The real uncertainty is the small-radius tail of the grain size distribution which may still be sufficient to reduce the electron and ion fractions at a few scale heights despite the tendency of grains to aggregate and settle towards the mid-plane. 

The robust point is that the relative magnitudes of the diffusivities are independent of the ionisation fraction, being controlled by the ratio $B/n_\mathrm{H}$, and that Hall diffusion dominates the other mechanisms over much of protoplanetary discs \citep{WN99,BT01,SS02a}. 
As X-rays and cosmic rays both penetrate to the Hall-dominated region just below the disc surface, it is absolutely essential that Hall diffusion be included in modelling the extent of MRI-driven turbulence in protoplanetary discs.  The real uncertainty is that, notwithstanding the pioneering effort of \citet{SS02a,SS02b}, the saturation of MRI-driven turbulence in the Hall-dominated regime 
 as yet remains unexplored.  

\section{Summary \& Conclusions}
\label{sec:summary}

In this paper we re-examined the role of Hall diffusion in suppressing or enhancing MRI-driven magnetic turbulence in  Keplerian discs.  

We first undertook a local, linear analysis of the magnetorotational instability, for simplicity restricting our attention to a vertical, weak magnetic field subject to axisymmetric perturbations with a purely vertical wave vector. While this is not new, our approach and presentation differed from previous analyses \citep{W99,BT01} in two critical ways.  First, we characterised the departure from ideal MHD using magnetic diffusivities rather than conductivities or characteristic frequencies, allowing us to make a clearer connection with the $\E\cross\B$ field-line drift implicit in the diffusive MHD induction equation.  Second, we presented a clear perspective of the dependence on the Hall and Pedersen (Ohm+ambipolar) diffusivities by examining how the properties of the instability vary over a Hall-Pedersen diffusivity plane (see Fig.~\ref{fig:locii}). This clearly delineates when the range of unstable wave numbers is finite or infinite, and whether the fastest growth occurs for finite wave number or asymptotically as $k\rightarrow\infty$.  We also discussed the limiting forms of the dispersion relation, making a connection with the diffusive plane-parallel shear instabilities of \cite{K08}.

Next, we reviewed the alternative parameterisations of non-ideal MHD that have appeared in the literature, and emphasised that existing simulations of the nonlinear development and saturation of the instability in the Hall-Ohm case \citep{SS02a,SS02b} are consistent with expectations based on the simple linear analysis and have not yet probed the Hall-dominated regime characteristic of protoplanetary discs. 

Finally, we illustrated the critical effect of Hall diffusion on the size of dead zones in \ppd s by applying  a local criterion for growth of the \mri \ to a simple model of minimum-mass solar nebula at 1\,au , including x-ray and cosmic-ray ionisation and a population of 1\,\micron\ grains.

Our key results can be summarised as follows.
\begin{enumerate}
    \item  The radial diffusion of perturbed field lines through the fluid is directly related to the criterion for marginal stability.  The MRI is suppressed if the radial drift of field lines  against the infall of the fluid is sufficient to restore the field to its equilibrium radial position.
    
    \item The behaviour of the MRI for our adopted geometry is determined by the dimensionless Pedersen and Hall diffusivities $\etaP \Omega/v_A^2$ and  $s\etaH\Omega/v_A^2$, where $s = \mathrm{sign}(B_z)$.  For $s\etaH<-2 v_A^2/\Omega$ Hall  diffusion suppresses the MRI irrespective of the value of $\etaP$.   For $s\etaH>-2 v_A^2/\Omega$ the $\etaP-s\etaH$ half plane is divided into three regions in which either (I) there is a finite range of unstable wave numbers with a fastest growing mode; (II) instability for all  wave numbers, with a unique fastest growing mode and a slower asymptotic growth rate as $k\rightarrow\infty$, or (III) instability at all wave numbers with fastest growth asymptotically achieved as $k\rightarrow\infty$ (see Fig.~\ref{fig:locii}). 
    
    \item For fixed $\etaP$, the maximum growth rate increases with increasing $s\etaH$, from 0 at $s\etaH\Omega/v_A^2 = -2$ to the maximum 0.75\,$\Omega$ as $s\etaH \rightarrow +\infty$. For fixed $\etaH$ with $s\etaH>-2 v_A^2  \Omega$, the growth rate is a maximum for $\etaP=0$ and declines as $\etaP\rightarrow\infty$ (see Fig.~\ref{fig:contours}).
    
    \item In the highly diffusive limit the instability reduces to the Hall-diffusion-driven instability in plane-parallel shear flow discussed by \cite{K08}.  Diffusion is so severe in this limit that the perturbations in the fluid velocity do not affect the field evolution, which is driven purely by diffusion and  Keplerian shear.  Our restriction to vertical initial fields and perturbation wave numbers enabled us to extend the results of \cite{K08} to include the damping by Pedersen diffusion and to derive pleasant analytic expressions for the upper cutoff to the unstable wavenumber range and the growth rate and wave number of the most unstable mode (see eqs \ref{eq:kc-shear}, \ref{eq:nu0-shear}, and \ref{eq:k0-shear}, respectively, and also Fig.~\ref{fig:nu-k-diffusion}).
    
    \item We argued that simulations of MRI-driven MHD turbulence in the presence of Hall and Ohm diffusion \citep{SS02a,SS02b} have not yet probed the ``deep'' Hall regime $s\etaH \ga \etaP \ga v_A^2/\Omega$, where the linear analysis suggests that Hall diffusion allows the instability to proceed when it otherwise would not.
    
    \item We found that at 1\,au in the minimum-mass solar nebula, Hall diffusion changes the magnetically active column density by an order of magnitude.  This change is either an increase or decrease depending on whether B is parallel or antiparallel to the rotation axis, respectively.  Hall diffusion likely plays a critical role in determining the radial extent of dead zones and the thickness of magnetically active layers in protoplanetary discs, and estimates based on damping by Ohm diffusion are probably wildly inaccurate.
    
\end{enumerate}
The simplifications adopted in our analysis engender three significant uncertainties in our conclusions that are worth some final discussion.

First, the restriction to the stability of vertical magnetic fields to perturbations with vertical wave vectors does not capture the destabilisation by ambipolar diffusion that arises when toroidal field and radial wave vector components are also present \citep{KB04,D04}. However, we have captured the analogous Hall diffusion-driven destabilisation of the MRI, and as Hall diffusion typically dominates this restriction is unlikely to have a great impact.

Second, while the use of the linear analysis to predict the boundary of the manifestly nonlinear active region, appears to be justified for the Ohm case \citep{TS08}, it is not known whether this applies in the Hall-dominated regime that we tout here.

Finally, we caution that the MRI may simply be irrelevant in protoplanetary discs.  A minor population of small dust grains would remove so many electrons from the gas phase that the MRI-active column density becomes so small as to be irrelevant.  Instead, magnetic activity -- if any -- may be due to fields lying within the disc and varying on length scales of order $r$ \citep{TS08} or a strong poloidal magnetic field brought in during the formation of the disc.

\section{Acknowledgements}
\label{sec:acknowledgments}
We are indebted to Catherine Braiding, Matthew Kunz, BP Pandey, Takayoshi
Sano, and Jim Stone for stimulating discussion and suggestions, and are
grateful for the hospitality provided by the Isaac Newton Institute for
Mathematical Sciences at Cambridge University and the School of Physics at
the University of Sydney, where some of this work was conducted.  This
research was supported by the Australian Research Council through Discovery
Project grant DP0881066, a Visiting Fellowship from the Isaac Newton Institute at the University of Cambridge, and an Outside Studies
Program grant from Macquarie University.

\bibliographystyle{mn2e}
\bibliography{refs}

\appendix

\section{Formulation}
\label{app:formulation}

Here we give the MHD equations describing a non-self-gravitating disc orbiting a point mass $M$.  We adopt cylindrical coordinates $(r,\phi,z)$ centred on $M$ with the disc mid-plane corresponding to $z=0$ and the disc angular velocity vector parallel to the z-axis.  We write the fluid equations in the inertial ``laboratory frame'' with the fluid velocity written as $\vv+\vvk$, so that $\vv$ denotes the \emph{departure} of the flow from the Keplerian velocity field
\begin{equation}
    \vvk = \sqrt{\frac{GM}{r}} \, \phih \,,
    \label{eq:vvk}
\end{equation}
which we eliminate in favour of the  Keplerian angular velocity 
\begin{equation}
    \mathbf{\Omega} = \frac{v_{\rm K}}{r} \, \zh
    \label{eq:Omega}
\end{equation}
using the identities $\div \vvk = 0$ and $\curl\vvk = \half \mathbf{\Omega}$.  This approach avoids the complexity that would be introduced by transforming to a local non-inertial frame or adopting a shearing-sheet approximation, neither of which are needed in our simple linear analysis in the next section.

The continuity equation becomes
\begin{equation}
\Delt{\rho} + \div (\rho\vv ) = 0 \,.    \label{eq:cty}
\end{equation}
The advective term containing $\Omega$ corresponds to $(\vvk\cdot\grad)\rho$ and represents the azimuthal advection associated with Keplerian rotation -- analogous terms will arise in the momentum and induction equations.  The momentum equation,
\begin{eqnarray}
\lefteqn{\Delt{\vv} + (\vv\cdot\grad)\vv - 2\Omega v_\phi\rh + \half\Omega v_r \phih 
=} \hspace{1cm} \nonumber\\ 
 && r\Omega^2 \rh - \grad\Phi -  \frac{1}{\rho}\grad P + 
 \frac{\J\cross\B}{c\rho}\,,    \label{eq:momentum}
\end{eqnarray}
also picks up $\Omega$\,-bearing terms that account for centripetal acceleration and the angular momentum loss implicit in radial motion if the azimuthal speed is Keplerian. The gravitational potential is
\begin{equation}
    \Phi = -\frac{GM}{\sqrt{r^2+z^2}}\,,
    \label{eq:Phi}
\end{equation}
and we adopt an isothermal equation of state:
\begin{equation}
    P = \rho c_{\rm s}^2 \,,
    \label{eq:eos}
\end{equation}
where $c_{\rm s}$ is the isothermal sound speed\footnote{This will not really be needed as the mode of interest does not involve pressure fluctuations.}.  The current density satisfies Amp\`eres Law:
\begin{equation}
    \J = \frac{c}{4\pi}\curl\B\,,
    \label{eq:J}
\end{equation}
and the magnetic field is, of course, solenoidal,
\begin{equation}
    \div\B=0\,,
    \label{eq:divB}
\end{equation}
and evolves according to the induction equation (\ref{eq:induction-v2}).
The qualitative discussion of the effect of field diffusion and the \mri\ in \S\ref{sec:diffusion} suggests that it will be useful to recast the induction equation in a form that makes explicit the drift of the magnetic field through the fluid, i.e.
\begin{equation}
    \Delt{\B} + \thalf\Omega B_r \phih =  
    \curl\left[(\vv+\vv_{\rm B})\cross\B \
    -\etaO(\curl\B)_\parallel\right]
    \,,
    \label{eq:induction-v2}
\end{equation}
where
\begin{equation}
    \vv_{\rm B} =  \etaP\,\frac{(\curl\B)_\perp\cross\Bh}{B} 
    \,-\,\etaH\,\frac{(\curl\B)_\perp}{B} \,.
    \label{eq:vB}
\end{equation}
Note that when $\J_\parallel$ vanishes, ambipolar and Ohm diffusion behave identically, appearing only together in sum as $\etaP$. This is the case in the linear analysis of the MRI presented here.

Note that the local dissipation rate associated with magnetic diffusion is
\begin{equation}
    \J\cdot\E' = \frac{4\pi}{c^2} \left[ \etaO\J_\parallel^2 + 
    \etaP\J_\perp^2 \right]
    \label{eq:JdotE}
\end{equation}
and so, as is well known, Hall diffusion has no associated dissipation. When $\J_\parallel$ vanishes, ambipolar and Ohm diffusion behave identically, appearing only together in sum as $\etaP$; in particular this is the case in the linear analysis of the MRI presented here.

Despite the appearance of $\Omega$-bearing terms, the equations we have derived apply to an arbitrary fluid flow -- we have not yet assumed that the flow is close to Keplerian, or even disc-like, but have simply written the fluid velocity in the MHD equations as $\vv+\vvk$. Of course, this form of the equations is only really useful for nearly Keplerian flows with $|\vv|\ll v_{\rm K}$. In the next section we shall specialise to a near-Keplerian disc equilibrium state with a vertical magnetic field, and consider perturbations with a vertical wave vector.

\section{Linear Analysis}
\label{app:analysis}

We consider a small region of an axisymmetric, geometrically thin and
nearly Keplerian disc, threaded by a vertical magnetic field, with sound
and Alfv\'en speeds ($c_{\rm s}$ and $v_{\rm A}$) at the  mid-plane that are
both small compared to the local  Keplerian speed $v_{\rm K}$.  We assume
that radial gradients are on the scale of $r$ and neglect vertical
stratification of the initial equilibrium state, so that our analysis only
holds near the  mid-plane, at heights $z\ll c_s /\Omega$.  Then, we may
neglect the term $ r\Omega^2 \rh - \grad\Phi$ in eq (\ref{eq:momentum}),
and the remaining radial and azimuthal derivatives appearing in eqs
(\ref{eq:cty})--(\ref{eq:induction-v2}), leaving only partial derivatives in
$t$ and $z$.

The initial state has $\vv=0$ (i.e.\ is in Keplerian rotation) with a
uniform density, pressure, and vertical magnetic field $\B=sB\zh$
(where $s=\pm1$).  We linearise the equations around this state and
seek solutions for axisymmetric perturbations of the form $\exp(\nu t
- ikz)$.  The equations for perturbations in density, pressure and 
$v_z$ form a separate system that describes vertically-propagating  
sound waves.  The system of linear equations in the remaining 
perturbations involve fluctuations in the $r$ and $\phi$ components of 
$\B$, $\vv$ and $\vv_{\rm B}$.  We treat these as two-component vectors in 
the equations below.

In the following equations, we recast physical quantities in 
dimensionless form by adopting $\Omega$ and $v_{\rm A}$ as the units of frequency 
and velocity, respectively.  For the remainder of this section, $\etaP$, 
$s\etaH$, $\nu$ and $k$  denote
$\etaP\Omega/v_{\rm A}^2$, 
$ \etaH\Omega/v_{\rm A}^2$, 
$\nu/\Omega$ and $k v_{\rm A}/\Omega$, respectively.  
Then the linearised momentum, field-line drift, and induction equations are
\begin{equation}
    \frac{\dv}{v_{\rm A}} = \frac{-ik}{(1+\nu^2)}\mat{\nu}{2}{-\half}{\nu} 
    \frac{\dB}{B}\,,
    \label{eq:dv}
\end{equation}
\begin{equation}
    \frac{\dv_B}{v_{\rm A}} = -ik\mat{\etaP}{s\etaH}{-s\etaH}{\etaP} 
    \frac{\dB}{B}\,,
    \label{eq:dvB}
\end{equation}
and
\begin{equation}
    \mat{\nu}{0}{\thalf}{\nu}\frac{\dB}{B} +
    ik \left(\frac{\dv}{v_{\rm A}} + \frac{\dv_B}{v_{\rm A}}\right) = 0\,.
    \label{eq:lin-induction}
\end{equation}
Using eqs (\ref{eq:dv}) and (\ref{eq:dvB}) to 
substitute for $\dv$ and $\dv_{\rm B}$ in eq. 
(\ref{eq:lin-induction}) yields 
the relationship between the 
components of $\dB$,
\begin{equation}
    \delta\!B_r = -\frac{s\etaH + 2/(1+\nu^2)}{\etaP+\nu A} 
    \,\delta\!B_\phi
    \label{eq:dBr}
\end{equation}
where
\begin{equation}
A = \frac{1}{1+\nu^2} + \frac{1}{k^2} \,,
\end{equation}
and the dispersion relation
\begin{equation}
	a k^4 + b k^2 + c = 0 \,
	\label{eq:dispersion_relation}
\end{equation}
where
\begin{eqnarray}
  a &=& (1 + \nu^2)(\etaP^2+\etaH^2) + {\textstyle \frac{5}{2}}s\etaH
+ 2\nu \etaP  + 1 \,, \label{eq:acoeff}\\[6pt]
  b &=&   (1 +\nu^2) \left( 2\nu\etaP - 
	{\textstyle \frac{3}{2}} s\etaH \right) + 2\nu^2 - 3 \,,
	\label{eq:bcoeff} \\[6pt]
  c &=& \nu^2(1+\nu^2) \,.
	\label{eq:ccoeff}
\end{eqnarray}
\citep{W99}\footnote{There is a typographical 
error in eq (14) of \citet{W99} -- the RHS should be preceded by a 
minus sign.}. While this form of the dispersion relation is convenient for determining the run of $\nu$ with $k$ for a given choice of the diffusivity components, it does not provide an overview of the dependence of the characteristic properties of the instability -- the range of unstable wave numbers, maximum growth rate and corresponding wave number -- on magnetic diffusion.  To this end, we recast the dispersion relation into a form that emphasises its dependence on the diffusivities $\etaH$ and $\etaP$: 
\begin{equation}
    \left(s\etaH + 
    \frac{2}{1+\nu^2}-\frac{3A}{4}\right)^{2} + 
    \Big(\etaP + \nu A \Big)^2  = \left(\frac{3A}{4}\right)^2 
    \,.
    \label{eq:disp-circle}
\end{equation}
Conveniently, a given growth rate and wave number corresponds to a circular
locus in the $s\etaH - \etaP$ plane, centred at $(3A/4 - 2/(1+\nu^2), -\nu
A)$ with radius $3A/4$, making this form amenable to graphical analysis.

Using this expression it is straightforward to show that the maximum
possible growth rate is $\nu = 3/4$ (in units of $\Omega$)
and that this occurs only when $\etaP = 0$.  To see
this, first note that $\etaP\geq 0$, and $A>0$.  Then, for $\nu\geq0$, the
LHS of eq (\ref{eq:disp-circle}) is $\geq \nu^2 A^2$, with equality
attained only if $\etaP = 0$ and $s\etaH = 3A/4-2/(1+\nu^2)$.  This
statement must also hold for the RHS of (\ref{eq:disp-circle}), so we
conclude that $\nu^2 \leq 9/16$ with equality only holding when $\etaP = 0$
and $s\etaH = 3/4k^2-5/4$.

Consider now the dependence of the instability on the  Pedersen and
Hall diffusivities $\etaP$ and $\etaH$ (see Fig.\ \ref{fig:locii}).  
First, note that the ideal-MHD limit holds at the origin (i.~e.~$\etaH=\etaP=0)$, the Hall MHD limit holds along the horizontal axis ($\etaP=0$) and the Ohm or ambipolar diffusion limits hold along the vertical line ($\etaH=0$).  Recall also that only the half plane $\etaP \geq 0$ is physically relevant. Inspection of the dispersion relation, eq.~(\ref{eq:disp-circle}), shows that all modes are stable for $s\etaH\leq-2$, and in the unstable region ($s\etaH>-2$) there is, at most, a single unstable mode for a given choice of wave number $k$.  
The run of growth rate with wave number, i.e.  $\nu(k)$, for a particular choice of $\etaH$ and $\etaP$, can be found by directly solving (\ref{eq:dispersion_relation}) for $\nu$.  In practice, it is easier to choose $\nu$ and solve the quadratic equation (\ref{eq:dispersion_relation}) for $k^2$.  It turns out that there are three distinct forms of the resulting $\nu(k)$ curve, corresponding to regions I, II and III in the $s\etaH$-$\etaP$ plane, as illustrated in Fig.\ \ref{fig:locii}. 

In region I, which lies outside the semi-circular locus
\begin{equation}
    \etaP^2 + (s\etaH + 5/4)^2 = 9/16 \,,
    \label{eq:kcinf-locus}
\end{equation}
the range of unstable
wave numbers extends from $k=0$ up to a maximum value
\begin{equation}
    k_c = \left[ \frac{3}{2}\, \frac{s\etaH+2}{\etaP^2 + (s\etaH + 
    5/4)^2 - 9/16}\right]^{1/2} \,
    \label{eq:kc}
\end{equation}
found by setting $\nu=0$
and $k=k_c$ in the dispersion relation (\ref{eq:disp-circle}).
The form of $\nu(k)$ in this region is illustrated in inset I of Fig.  \ref{fig:locii}.  The maximum growth
rate $\nu_0$ corresponds to a repeated root for $k^2$ in the quadratic
(\ref{eq:dispersion_relation}), so it can be found by setting the
discriminant to zero.  This yields
\begin{equation}
     s\etaH =\frac{24\,\nu_0}{9-16\nu_0^2}\,\etaP - \frac{2}{1+\nu_0^2}\,,
    \label{eq:nu0-1}
\end{equation}
so contours of constant $\nu_0$ in region I are straight lines.
The corresponding wave number, $k_0$, is then given by the ratio
$-2c/b$, or alternatively
$-b/2a$, obtained from eqs (\ref{eq:acoeff})--(\ref{eq:ccoeff}) with $\nu$ set to 
$\nu_0$, i.e.
\begin{equation}
  k_0 = \left[ \frac{-2\nu_0^2(1+\nu_0^2) }
  { (1 +\nu_0^2) \left( 2\nu_0\etaP - {\textstyle \frac{3}{2}} s\etaH 
  \right) + 2\nu_0^2 - 3}\right]^{1/2}\,.
    \label{eq:k0}
\end{equation}
Region I encompasses the limits of ideal-MHD, the Ohm and/or ambipolar 
diffusion limit, as well as the Hall limit (i.e.~$\etaP=0$) for $s\etaH>-0.5$.
Elsewhere, i.e.  within the semicircle bounded by  (\ref{eq:kcinf-locus}) and
the $s\etaH$-axis ($\etaP=0$), all wave numbers are unstable.  The interior of
the semicircle is further subdivided into regions II and III, depending on
whether the maximum growth rate is attained at a finite wave number or
asymptotically as $k\rightarrow\infty$, as illustrated in insets II and III in
Fig.~\ref{fig:locii}, respectively.

In region II, $\nu_0$ and $k_0$ still satisfy eqs (\ref{eq:nu0-1}) and
(\ref{eq:k0}) just as in Region I. The inner boundary of region II occurs where
$k_0$ just becomes infinite, ie.  $\nu_0$ satisfies eq (\ref{eq:nu0-1}) 
and simultaneously the denominator of eq (\ref{eq:k0}) is zero.  These two 
conditions yield a parametric solution
for the locus separating regions II and III,
\begin{eqnarray}
    s\etaH &=& -\frac{2(9+4\nu_0^2)}{(1+\nu_0^2)(9+16\nu_0^2)}\nonumber\\ 
        \label{eq:k0inf-locus}\\
    \etaP &=& \frac{\nu_0(9-16\nu_0^2)}{(1+\nu_0^2)(9+16\nu_0^2)}\,,  \nonumber
\end{eqnarray}
which traces out an arc from $(\etaP,s\etaH) = (0,-2)$ to 
$(0,-\frac{4}{5})$ as $\nu_0$ runs from $0$ to $3/4$ (the blue locus 
in Fig. \ref{fig:locii}).  

In region III, all wave numbers are unstable, with the fastest growth 
occuring in the limit $k\rightarrow\infty$, yielding from 
(\ref{eq:disp-circle}):
\begin{equation}
    \left(s\etaH + 
    \frac{5/4}{1+\nu_0^2}\right)^{2} + 
    \left(\etaP + \frac{\nu_0}{1+\nu_0^2}\right)^2 = 
    \left(\frac{3/4}{1+\nu_0^2}\right)^2 \,.
    \label{eq:nu0-2}
\end{equation}
Thus in region III, contours of constant $\nu_0$ trace out segments of non-concentric 
circles running between the $s\etaH$-axis and the boundary with region 
II.

Having delineated these three regions, we now consider how the critical
wave number $k_c$, fastest growth rate $\nu_0$ and corresponding wave
number $k_0$ vary across the entire $\etaP$--$s\etaH$ plane.
 Contours of constant $k_c$ 
are semicircles, as plotted in  Fig.\ \ref{fig:kcrit}.
While the range of unstable wave numbers is reduced for large values of  
$s\etaH$ and $\etaP$, as one might expect, the range is not maximised in the
ideal limit (ie.  at the origin) but in regions II and III, bounded by the
$k_c=\infty$ contour.

Turning now to the fastest growing modes, the maximum growth rate is given either by (\ref{eq:nu0-1})
in regions I and II or by (\ref{eq:nu0-2}) in region III, and the 
corresponding contours are plotted
in Fig.\ \ref{fig:contours}.    
The growth rate
increases clockwise, from $0\,\Omega$ along the
vertical line $s\etaH=-2$ up to $0.75\,\Omega$ for the horizontal line $\etaP =
0$ for $s\etaH>-4/5$.  In the absence of Hall diffusion, the maximum growth
rate $\nu_0$ declines with increasing 
(Ohm and/or ambipolar)
diffusivity
(e.g. moving vertically upwards through the $s\etaH = 0$ point in the horizontal axis), 
with $\nu_0 \approx
\textstyle{\frac{3}{4}\etaP^{-1}}$ for $\etaP\gg 1$.  The most
important effect of Hall diffusion, apparent from Fig.\
\ref{fig:contours}, is that 
\emph{the growth rate of the MRI exceeds
$0.3\Omega$ for $s\etaH\ga \etaP$, even for arbitrarily large $\etaP$}.
More generally, the addition of Hall diffusion at fixed $\etaP$
increases the growth rate if $s\etaH>0$ and decreases it when
$s\etaH<0$.  For large values of $\etaP$, eq (\ref{eq:nu0-1}) shows
that $s\etaH/\etaP \approx 24\nu_0/(9-16\nu_0^2)$.  It is this fact
that has the potential to modify the extent of dead zones in protoplanetary discs, as we explore later in \S \ref{sec:ppds}.

The wave number of the fastest growing mode (blue contours in Fig.\
\ref{fig:contours}) 
decreases as the diffusivity is increased.  Again, the
contours are not arranged so that the highest wave numbers occur in the
ideal-MHD limit, but to the $s\etaH<0$ side, within the boundary between
regions II and III (traced by the $k_0=\infty$ contour).

Overall, these patterns place the ideal, Ohm (or ambipolar) and Hall
regimes in context, and for the first time we see an overview of the effect
of magnetic diffusivity on the linear MRI. In particular, there is nothing
special about the Ohm/ambipolar limit, e.g.\  the behaviour of the
instability in the presence of diffusion is not qualitatively different for
$s\etaH=2$ vs $\etaH=0$.  Even the ideal-MHD limit does not stand apart as remarkable,
although it still holds a special place conceptually because flux 
freezing holds and it is easier to think about.   What does stand out is the 
part of the plane in the lower left, regions II and III, characterised by 
high wave numbers and the spraying out of the growth contours.

Overall we note that increasing $\etaP$ decreases the maximum growth rate
and the characteristic wave numbers, whereas increasing $s\etaH$ above $-2$
increases the maximum growth rate and may either increase (when
$s\etaH+1\la\etaP$) or decrease (when $s\etaH+1\ga\etaP$) the corresponding
wave number.

\section{Limiting cases and diffusive instabilities}
\label{app:limits}
In this section we consider the interesting limiting cases of our analysis.  As a preliminary, we substitute equations (\ref{eq:dv}) and (\ref{eq:dvB}) into (\ref{eq:lin-induction}) to express the linearised induction equation in the form
\begin{eqnarray}
\Bigg[ \; \mat{\nu}{0}{\thalf}{\nu}
& + & \frac{k^2}{1+\nu^2} \mat{\nu}{2}{-\half}{\nu} \nonumber \\
& + & \; k^2 \mat{\etaP}{s\etaH}{-s\etaH}{\etaP}\; \; \Bigg] \; \dB \; = \; 0 \,.
\label{eq:dB-total}
\end{eqnarray}
The three terms in this expression represent the effects of Keplerian shear, fluid displacement, and magnetic diffusion on the magnetic field perturbations.   Each of three limiting cases can be obtained by neglecting one of these terms.  To obtain the criteria for each limit, we note that $\nu\leq\frac{3}{4}$, and so the three terms have order of magnitude 1, $k^2$, and $k^2\eta_\perp$, respectively.  The three cases of interest are as follows.

\subparagraph{Ideal MHD} When $k^2\eta_\perp \ll k^2\sim 1$, i.e. $\eta_\perp \ll 1$, $k^2\sim 1$, the third term is negligible, and field evolution is determined by shearing of the field and the response of the fluid to magnetic stresses.  This limit applies in the neighbourhood of the origin in Fig.~\ref{fig:contours}, and (of course) recovers the results of \cite{BH91} limited to a vertical field and wave vector and neglecting buoyancy.

\subparagraph{Cyclotron limit} The first term is negligible when $k^2\eta_\perp \sim k^2 \gg 1$ (i.e. $\eta_\perp \sim 1$ and $k^2\gg 1$).  In this limit the important effects are advection by the fluid displacement and magnetic diffusion.  Generation of $B_\phi$ from $B_r$ by the Keplerian shear flow is negligible in this limit, but Coriolis and centripetal acceleration still play a crucial role through the dynamics of the fluid which enters via the appearance of the off-diagonal matrix elements in the 2nd term in eq (\ref{eq:dB-total}).

This short-wavelength, low-frequency limit corresponds to the cyclotron mode of the magnetised fluid, which has frequency
\begin{equation}
    \omega_H = \frac{v_A^2}{|\etaH|} =  \frac{eB}{m_i c}\,\frac{\rho_i}{\rho}\,,
    \label{eqn:omegaH}
\end{equation}
where the 2nd form applies for a simple ion-electron-neutral plasma \citep{WN99,PW08}
This mode is able to couple effectively to the Keplerian rotation as long as the sense of circular polarisation of the mode matches the epicyclic motion, i.e. as long as $B_z\etaH < 0$.   The other short-wavelength mode, the high-frequency whistler ($\omega = k^2 v_A^2/\omega_H$)\footnote{In dimensionless form, $\omega_H = 1/|\etaH|$ and the whistler mode has frequency $\omega = k^2/|\etaH|$.}, is unable to couple effectively to the rotation \citep{W99}.

This limit applies for $k\rightarrow\infty$ in regions II and III of Fig.~\ref{fig:locii}, where arbitrarily large wavenumbers are unstable.  The dispersion relation in this case is
obtained by letting $k\rightarrow\infty$ in (\ref{eq:dispersion_relation}), and reduces to
\begin{equation}
    (1 + \nu^2)(\etaP^2+\etaH^2) + {\textstyle \frac{5}{2}}s\etaH
+ 2\nu \etaP  + 1 =0 \,,
\end{equation}
i.e.~$a(\nu)=0$ with $a$ given by eq (\ref{eq:acoeff}).  The lack of any $k$ dependence in this regime occurs because both the magnetic diffusion and the magnetic stresses on the fluid (which are responsible for the fluid displacement) scale as $k^2$.

\subparagraph{Diffusive limit}
Finally, the 2nd term in (\ref{eq:dvB}) may be neglected when $k^2\eta_\perp\sim 1 \gg k^2$ (i.e. $\eta_\perp \gg 1$, $k^2\sim 1/\eta_\perp$).  In this case instability relies on the keplerian shear flow generating $B_\phi$ from $B_r$, and the tendency of Hall diffusion to convert $B_\phi$ back into $B_r$.  This brings the potential destabilising effect of Hall diffusion in shear flows to the fore and shows that it is quite independent of rotational effects -- i.e.~the Coriolis and centripetal acceleration -- that drive the MRI
\citep{K08}.  Here our simplified geometry and large diffusion limit allow us to find simple analytic expressions for the growth rate as a function of $k$ for arbitrary diffusivity.

Eqs (\ref{eq:dv}) and (\ref{eq:dvB}) show that $|\dv|\ll|\dv_{\rm B}|$, so that the $\dv$ term can be neglected in the linearised induction equation (\ref{eq:lin-induction}), yielding:
\begin{equation}
 \mat{\nu}{0}{\thalf}{\nu}\frac{\dB}{B} +
    k^2\mat{\etaP}{s\etaH}{\!\!-s\etaH}{\etaP} 
    \frac{\dB}{B} = 0\,,
    \label{eq:lin-induction-diffn}
\end{equation}
with $\dv$ still given by eq (\ref{eq:dv}).  In this limit the field evolves in response to shear and diffusion, without significant feedback from the perturbations that it induces in the fluid flow. The 3/2-bearing term ($\equiv -d\ln\Omega/d\ln r$) is the only manifestation of the Keplerian rotation law and cylindrical geometry in this equation\footnote{The velocity perturbations, however, are still affected by rotation (see eq \ref{eq:dv}) but do not themselves feed back on the field evolution.}.  To generalise this expression we replace $\thalf\Omega$ by a characteristic shear frequency $v'$ that can be thought of either as accommodating different rotation laws in cylindrical geometry, or represents plane-parallel shear in a cartesian geometry.  In the latter case $r$ and $\phi$ are mapped to $-x$ and $-y$ with a local velocity field $\bm{v} = v'\, x \, \hat{\bm{y}}$; the $z$-axis is unchanged.  Without loss of generality we may assume that $v'>0$, and as before, we write $\B=sB\zh$ with $s=\pm1$.

Upon replacing the $\thalf$ term in (\ref{eq:lin-induction-diffn}) by $v'$, we obtain the dispersion relation in dimensional form: 
\begin{equation}
    \left(k^2\etaP+\nu\right)^2 + \left(k^2s\etaH-v'/2\right)^2 = (v'/2)^2 \,.
    \label{eq:disp-shear}
\end{equation}
This immediately shows that instability requires both $v'$ and $s \etaH$ to be nonzero and have the same sign, otherwise the 2nd term exceeds the right hand side and there are no acceptable solutions.  
The dispersion relation is easily solved for the growth rate
\begin{equation}
    \nu = \sqrt{k^2s\etaH(v' - k^2s\etaH)} - k^2\etaP\,.
    \label{eq:nu-shear}
\end{equation}
and the equilibrium is unstable to wave numbers running from 0 up to a 
cutoff
\begin{equation}
    k_c = \sqrt{\frac{s\etaH\,v'}{\eta_\perp^{\,\,2}}}\,,
    \label{eq:kc-shear}
\end{equation}
where, as usual, the perpendicular diffusivity is 
\begin{equation}
    \eta_\perp = \sqrt{\etaH^2+\etaP^2}\,.
    \label{eq:etaperp}
\end{equation}

To find the fastest growing mode we play the usual trick: write eq
(\ref{eq:disp-shear}) as a quadratic in $k^2$ and find $\nu_0$, the value of $\nu$
that gives zero discriminant.  This means that there is only one corresponding wavenumber $k_0^2$
and therefore the point $(k_0,\nu_0)$ lies on the peak of the
curve $\nu(k)$.  The wave number is found from the quadratic, which is
easily solved when the discriminant vanishes.  Fortuitously, this procedure
yields simple expressions for the maximal growth rate,
\begin{equation}
    \nu_0 = \frac{s\etaH v'}{2(\eta_\perp+\etaP)} \,,
    \label{eq:nu0-shear}
\end{equation}
and corresponding wave number
\begin{equation}
    k_0 = \sqrt{\frac{s\etaH v'}{2\eta_\perp(\eta_\perp+\etaP)}}\,,
    \label{eq:k0-shear}
\end{equation}
where we have made use of the identity
\begin{equation}
    \frac{\eta_\perp-\etaP}{\etaH} = \frac{\etaH}{\eta_\perp+\etaP}
    \label{eq:etaperp-identity}
\end{equation}
to neatly avoid delicate subtractions when $|\etaH|\ll\etaP$.
Note that
\begin{equation}
    \nu_0 = k_0^2\eta_\perp \,,
    \label{eq:n0k0-shear}
\end{equation}
which we can then use in eq (\ref{eq:dBr}) along with $|\etaH|$, $\etaP\gg1$
to show that the relationship between the perturbed field components in the
fastest growing mode is
\begin{equation}
    \delta B_x = - \frac{s\etaH}{\etaP+\eta_\perp}\;\delta B_y \,.
    \label{eq:dbr-shear}
\end{equation}

The MRI proceeds in the ideal MHD limit because shear creates $B_\phi$ from $B_r$, $B_r B_\phi$ stresses on the fluid cause it to move radially inwards and outwards, creating more $B_r$ available to be converted to $B_\phi$ by the shear.  In the diffusive limit, the radial component of $\B$ is not generated by the response of the fluid to magnetic stresses, but because Hall diffusion converts $B_\phi$ to $B_r$. For more general field and wave vector configurations, ambipolar diffusion may play a similar role in assisting or suppressing the MRI, albeit hindered by dissipation \citep[Pandey \& Wardle, in prep]{KB04,D04,K08}.

\bsp
\label{lastpage}
\end{document}